\documentclass[12pt]{article}

\usepackage{amssymb,amsmath,mathrsfs,epstopdf,slashed,color}

\usepackage{graphicx}
\usepackage{hyperref}
\usepackage{cite}

\allowdisplaybreaks[1]

\makeatletter

\@addtoreset{equation}{section}
\makeatother

\newcommand{\vol}{\mathrm{vol}}

\title{\bf Population Preference \\in String Theory Landscape}
\author{}

\begin{document}

\begin{titlepage}

\setcounter{page}{0}
  
\begin{flushright}
 \small
 \normalsize
\end{flushright}

\vskip 3cm
\begin{center}
    
  {\Large \bf A Stringy Mechanism for A Small Cosmological Constant}
  
\vskip 2cm
  
{\large Yoske Sumitomo${}^1$ and S.-H. Henry Tye${}^{1,2}$}
 
 \vskip 0.7cm

 ${}^1$ Institute for Advanced Study, Hong Kong University of Science and Technology, Hong Kong\\
 ${}^2$ Laboratory for Elementary-Particle Physics, Cornell University, Ithaca, NY 14853, USA

 \vskip 0.4cm

Email: \href{mailto:yoske@ust.hk, iastye@ust.hk}{yoske at ust.hk, iastye at ust.hk}

\vskip 1.5cm
  
\abstract{\normalsize
 Based on the probability distributions of products of random variables, we propose a simple stringy mechanism that prefers the meta-stable vacua with a small cosmological constant. We state some relevant properties of the probability distributions of functions of random variables. We then illustrate the mechanism within the flux compactification models in Type IIB string theory. As a result of the stringy dynamics, we argue that the generic probability distribution for the meta-stable vacua typically peaks with a divergent behavior at the zero value of the cosmological constant. However, its suppression in the single modulus model studied here is modest.
  }
  
\vspace{4cm}
\begin{flushleft}
 \today
\end{flushleft}
 
\end{center}
\end{titlepage}

\setcounter{page}{1}
\setcounter{footnote}{0}

\tableofcontents

\parskip=5pt

\section{Introduction}

Recent cosmological data strongly suggests that our universe has been sitting at a vacuum state with an exponentially small positive vacuum energy density, or cosmological constant $\Lambda$.
The argument of Bousso and Polchinski \cite{Bousso:2000xa} (see also reviews \cite{Douglas:2006es,Bousso:2012dk}) suggests that string theory has so many possible vacua solutions that, in principle, it can have solutions that are consistent with such a very small positive $\Lambda$. However, their argument leaves open the most pressing question why string theory prefers a vacuum solution with such a very small $\Lambda$. Here we like to propose a plausible reason why this may happen within the context of string theory. The basic underlying physics is very simple. After pointing out the conditions under which a stringy solution with a small $\Lambda$ may be preferred, we illustrate with some examples in flux compactification in Type IIB theory.  

\begin{figure}
 \begin{center}
  \includegraphics[width=19em]{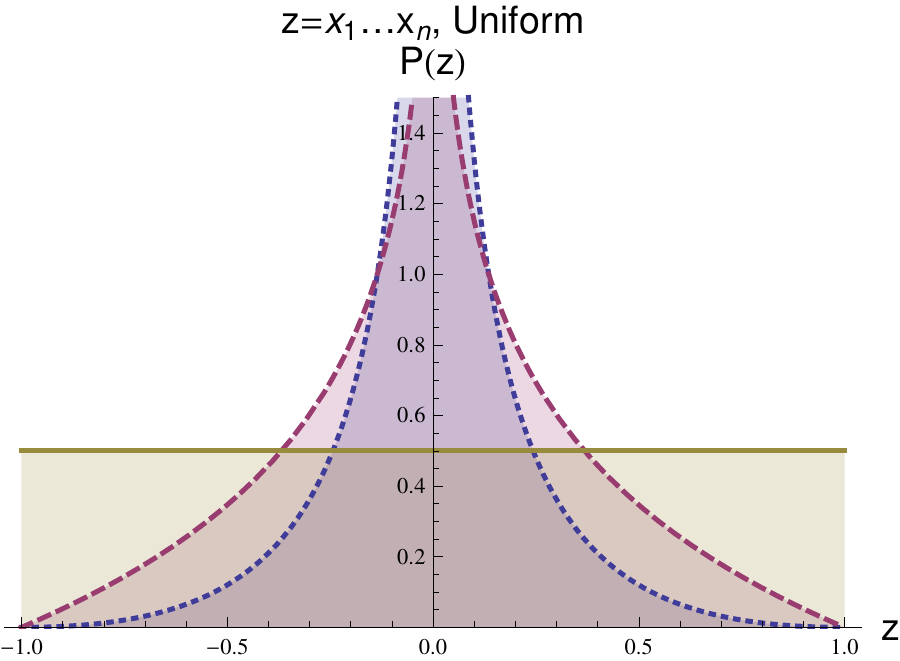}
  \hspace{1em}
  \includegraphics[width=19em]{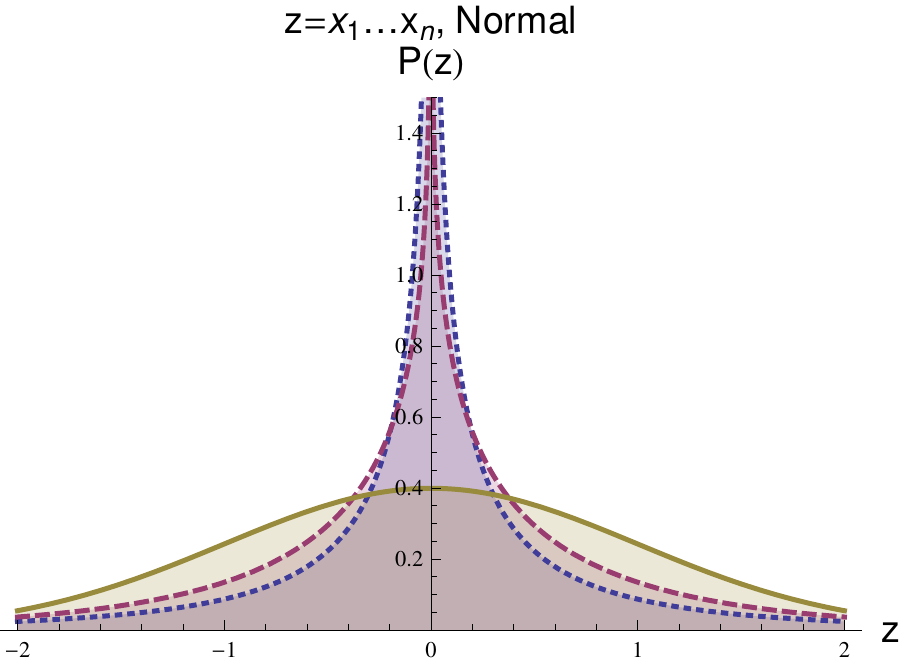}
 \end{center}
 \caption{\footnotesize Some examples of the product distribution of $z$ where each $x_j$ has an uniform ([-1,1]) distribution (on the left hand side) or a normal (Gaussian) distribution with variance $\sigma_j = 1$ (on the right hand side).
 The product distribution $P(z)$ is for $n =$ 1 ($z=x_1$, solid brown curve), $n=2$ ($z=x_1x_2$, red dashed curve), and $n=3$ ($z=x_1x_2x_3$, blue dotted curve), respectively. The curves on the left are given in (\ref{s11}).
 The curves on the right are given by the Meijer-G function (\ref{MeijerGn}), which reduces to the modified Bessel function of the second kind for $n=2$.}
 \label{fig:product distribution under uniform}
\end{figure}

The key idea follows from some simple well known properties of probability distributions for products of random variables. Consider a set of random variables $x_j \ (j=1, \cdots , n)$. As an example, if each random variable $x_j$ has a uniform probability distribution between $[-1, 1]$, then the probability distribution $P(z)$ for their product is given by
\begin{equation}
\label{s11}
{P} (z) = \frac{1}{2(n-1)!} \left(\ln {1 \over |z| } \right)^{n-1}
\end{equation}
for  $z=x_1 x_2 \cdots x_n$, where $-1 \le z \le 1$. Note that $P(z)$ peaks and in fact  diverges at $z=0$. 
This peaking property is insensitive to the particular probability distributions of $x_j$ as long as each distribution smoothly includes zero. For example, if each $x_j$ has a normal (Gaussian) distribution, then ${P} (z)$ still diverges like $\ln (1/|z|)^{n-1}$ at $z=0$.
For $z=x_1^n$, $P(z)$ diverges like $|z|^{-1 + 1/n}$ as $|z| \rightarrow 0$.
The probabilty distribution for $x_j$ and the product distributions for $z=x_1 x_2$ and $z= x_1x_2 x_3$ are illustrated in Figure \ref{fig:product distribution under uniform}.

To obtain a low energy effective theory in string theory, we have to compactify and stabilize the extra dimensions (see e.g. \cite{Douglas:2006es}).
A compactification is described by the compactified manifold's shape and sizes, which collectively are referred to as moduli. 
Different sets of moduli may be continuously morphed into each other as the geometry changes. The cosmic stringy landscape is described by these moduli. 
A specific flux compactification can be achieved via the introduction of a dozen or more types of $4$-form quantized fluxes \cite{Bousso:2000xa}, whose discrete values can easily range to many millions. This feature translates to moduli that can take numerous values. 
At low energies, the landscape may be described by an effective potential $V$, which is a function of all the light moduli. To simplify the analysis, we assume that most of the moduli except a few have already been stabilized at some high scales. They may be integrated out, so these stabilized moduli are now replaced by a set of parameters in the low energy effective potential $V$.  Now,  some of these parameters may take numerous values.  In sampling the landscape, we may treat some of the parameters as random variables taking values that smoothly include zero. 

In this paper, we study a particularly simple example in Type IIB flux compactification, namely,  the single modulus model in the K\"ahler  uplifting scenario \cite{Balasubramanian:2004uy,Westphal:2006tn,Rummel:2011cd,deAlwis:2011dp}  
in the large volume approximation \cite{Balasubramanian:2005zx}.
Consider the superpotential $W=W_0 - A e^{-x}$, where $W_0$ and $A$ are parameters and $x$ is a K\"ahler modulus. In the large volume approximation, the potential $V(x)$ may be reduced to 2 terms only. We are interested in a meta-stable vacuum (which may quantum tunnel to another vacuum state). Such a stable vacuum can exist at $x=x_m$. If so, $\Lambda$ at this stable vacuum (in a simplified version) turns out to have the form
\begin{equation}
\label{Leg}
\Lambda \simeq B W_0 A (x_m-x_0)
\end{equation}
where $B$ is a constant. In this particular model $x_0=5/2$.
Depending on the input parameters and the K\"ahler potential, we may tune $(x_m - x_0)$ to zero or to an exponentially small value. However, this is fine tuning. On the other hand, let us treat $W_0 \ge 0$ and $A \ge 0$ as random parameters.
Then, intuitively at least, the resulting probability distribution ${P} (\Lambda)$ for $\Lambda$ will naturally peak at $\Lambda=0$. That is, ${P} (\Lambda)$ in general diverges at $\Lambda=0$. 

Note that the above feature is where string theory differs from conventional field theory. In field theory, or supergravity, the parameters in $V$ take fixed values. So there is no distribution of $\Lambda$ to talk about, and the above peaking behavior is absent. It is the string landscape that provides the reason why some of the  parameters may be allowed to take multiple values \footnote{Of course one can endow a supergravity model with a collection of quantized 4-form fluxes and branes. We consider such models as string-motivated and the mechanism there as stringy as well.}. We shall make some plausible inputs so some of the parameters that enter into (\ref{Leg}) can take values that include zero. However, the specific peaking behavior is different with different plausible inputs, resulting in different levels of suppression of the expected value for $\Lambda$. So we see intuitively that the peaking behavior at $\Lambda=0$ is a stringy feature. It is the result of high scale string dynamics in flux compactification that is believed to be present.

Although the above peaking behavior captures some of the stringy features, it alone does not tell the whole story. The average magnitude of the naive $\Lambda$ (\ref{Leg}) simply follows from the product of the average magnitudes of the parameters emerging from string theory. In string theory, the actual $\Lambda$ is the value of a meta-stable minimum of $V$. The extremum ($\partial_k V=0$) and the stability (positivity of the Hessian matrix $\partial_k \partial_l V$) conditions impose a constraint on the ratio $W_0/A$. 
Apriori, this type of constraints may erase the peaking behavior coming from the above intuitive argument. In this case, if $W_0$, $A$ and $(x_m- x_0)$ smoothly include zero, we expect ${P}(\Lambda)$ to diverge like $(\ln\Lambda)^2$ (at $\Lambda=0$). However, as we shall see, the constraint cuts the power by one, to ${P} (\Lambda) \sim \ln (1/\Lambda)$. Treating other parameters in the model as random, $P(\Lambda)$ can even have a power-like divergence at $\Lambda=0$. However, with uniform distributions for the random parameters, the mechanism produces only a modest suppression of $\Lambda$. On the other hand, even with only 2 random parameters $W_0$ and $A$, they themselves may be non-trivial functions of the other stabilized moduli (at some higher scales) in flux compactification, so their distributions may be peaked already. 
If the probability distributions of $W_0$ and $A$ are themselves peaked at zero (which is very likely if we look at how they arise)\footnote{Although non-trivial solutions allow $W_0, A$ to take ranges of values that include zero, it is probably natural to simply have $W_0=A=0$, yielding supersymmetric Minkowski solutions. This suggests that the probability distributions of $W_0, A$ naturally peak at zero values.}, then ${P} (\Lambda)$ will be more sharply peaked at zero.
As the peak becomes sharper, a smaller $\Lambda$ becomes more likely.

As we just explained, the actual stringy mechanism for a small $\Lambda$ in the single modulus model in the K\"ahler  uplifting scenario \cite{Balasubramanian:2004uy,Westphal:2006tn,Rummel:2011cd} 
is a little involved due to the constraints. Under plausible conditions, we see that the peaking behavior for $\Lambda$ is modified but not erased by the constraints. Furthermore, the constraints tend to limit the ranges of the random parameters, so the resulting $\Lambda$ can be substantially smaller than the naive $\Lambda$.  
Even in such a very simple model, we already see that the preferred $\Lambda$ can be smaller than the typical values suggested by the string/Planck scale.
This reduction in the expected value of $\Lambda$ is the output of explicit moduli stabilization within string dynamics.
We believe that the preference for a small $\Lambda$ in string theory is a generic feature of the string dynamics in flux compactification.

In a scenario with a single modulus $\phi$, the presence of a $\phi$-independent term in $V$ typically erases the above peaking feature. That is, when a term that peaks at zero is added to an independent term, the peaking feature is usually erased. More generally, if any subset of moduli does not couple to the rest of the moduli (the above example has an empty subset), the above peaking behavior will generically be erased. So
we simply assert that such couplings among the moduli are necessary for the peaking phenomenon for a small $\Lambda$. 
In scenarios with more than one modulus, it is important that they are coupled to each other, directly or indirectly. Otherwise, $\Lambda$ ends up to be a sum of terms, typically erasing the peaking phenomenon. It is the extremum and the stability conditions that link the various terms in $V$ together. In this sense, the stringy dynamics is very important.

The smallness of $W_0, A$ also leads to a small modulus mass $m^2 \sim BW_0A$ in Planck unit.
The cosmological moduli problem \cite{Banks:1993en,deCarlos:1993jw} constrains the mass of moduli fields to be $m_\phi \gtrsim {\cal O}(10) \, {\rm TeV}$, which leads to (with peaked distributions for $W_0$ and $A$),
$\Lambda \gtrsim 10^{-30}$ (in Planck unit) in (\ref{Leg}).  
Here we see no particular reason (without fine tuning) why the factor $(x_m-x_0)$ in (\ref{Leg}) should be exponentially small. 
However, if the corresponding (to $(x_m-x_0)$) factor in a more non-trivial multi-moduli scenario can naturally be exponentially small, 
then the moduli masses may still be very small, but they may no longer be closely tied to the value of $\Lambda$.
It remains an open question whether (and under what conditions) this mechanism can naturally produce a $\Lambda$ as small as the one observed. It is plausible that there is a collection of mechanisms pushing $\Lambda$ down and the above mechanism is simply one of them.

Although the above examples suggest that small $\Lambda$s are preferred, they also seem to imply equal (or comparable) probabilities for both positive and negative values of $\Lambda$. For the preference of a positive value, we may invoke the history of our universe. Cosmological data suggests that our universe has gone through an inflationary epoch in its very early stage. So our universe most probably started with a relatively large positive vacuum energy density during the inflationary epoch; it subsequently moved down along a particular path in the ``cosmic landscape'' and eventually landed at the present small value.  Since the universe moved down from a positive large $\Lambda$, it passes the small positive $\Lambda$ region before reaching the negative $\Lambda$ region, so it is reasonable that our universe reached a meta-stable vacuum with a small positive $\Lambda$ before it has a chance to move further down to a negative $\Lambda$ vacuum. 

For a random variable that takes discrete values only, we shall assume that the spacings of the discrete values are small enough for us to ignore this fine point. In the evaluation of actual $\Lambda$ in a realistic model, we must make sure that the very small observed dark energy value can be reached. Invoking the Bousso-Polchinski argument, we believe this is a valid assumption. Since the model we are considering here is still rather crude, we shall not worry too much about this important issue in this paper. We do discuss the discreteness issue when it is directly relevant to the model we study.

The rest of the paper is organized as follows. We review and discuss the probability distribution of products (also ratios and sums) of random variables in Section \ref{sec:distr-prod-rati}, in preparation for their applications to examples in string theory. The key results are summarized in Table \ref{tab:asymptotics of product distribution}. In Section \ref{sec:bousso-polch-example}, we show how a preference for a small $\Lambda$ does not emerge in some toy models. In Section \ref{sec:large-volume-scen}, we discuss in some detail a Type IIB flux compactification model in the large volume scenario.
This model was studied analytically recently in \cite{Rummel:2011cd} (see also \cite{Westphal:2006tn,Parameswaran:2007kf}). We present and discuss the background necessary for our analysis.
In Section \ref{meaL}, we take the viewpoint explained above and see how the stringy mechanism for a small $\Lambda$ works in a single  modulus case. In particular, we discuss how the peaking behavior emerges. We consider different plausible scenarios and summarize the key features in Table \ref{tab:asymptote and vev of LVS}.  We end with some discussions and remarks in Section \ref{Remarks}.
Some details are relegated to Appendix \ref{sec:other-scen-param}.

\section{Distributions for products, ratios and sums of random variables \label{sec:distr-prod-rati}}

Let us review and discuss the probability distributions of various combinations of random variables, including the probability distribution of the product of random variables, namely, the {\it Product Distribution}, which will be a key ingredient governing the population behavior of the effective potential extrema for moduli stabilization models in string theory. Even if the distributions of random variables are simple, a product of these variables can become non-trivial.  In anticipation to our application in a flux compactification model in string theory, we also discuss the probability distributions for sums and ratios of random variables. The key results are easily derived and they are summarized in Table \ref{tab:asymptotics of product distribution}.

To begin, we like to specify the statistical properties of the random variables $x_j$.
We shall treat each $x_j$ as an independent, identically distributed (i.i.d.) variable drawn from some statistical distribution $\Omega_j$.
Different $x_j$ can have different $\Omega_j$.
For our discussion here, the precise choice of $\Omega_j$ is not important, provided that the higher moments of $\Omega_j$ are appropriately bounded.
(These higher moments are essentially the Mellin transforms that will play an important role here.)
However, it is important whether the zero value is inside $\Omega_j$ or not.
By definition, all distributions for $x_j$ and $y_i$ include zero smoothly.
For random variables that excludes zero, we shall call them $c_k$, with statistical distribution $\hat{\Omega}_k$.
For a random variable that takes discrete values only, we shall simply assume that the spacings of the discrete values are small enough for us to ignore this fine point.

\subsection{Product distribution}

The product distribution may be derived analytically if the original distribution is also known analytically.
For instance, if the distribution of each random variable $x_i$ obeys uniform distribution changing between 0 and 1, the properly normalized product distribution of $z = x_1 x_2 \cdots x_n$ is obtained by integrating out a delta-function \cite{Sakamoto:1943}:
\begin{equation}
 \begin{split}
  P(z) =& \int_0^1 \, dx_1 dx_2 \cdots dx_n \, \delta(x_1 x_2 \cdots x_n - z) = {1\over (n-1)!} \left(\ln {1\over z} \right)^{n-1}.
 \end{split}
 \label{product distribution for uniform}
\end{equation}
for $0 \le z \le 1$. If each $x_i$ obeys the uniform distribution in the range $[-1, 1]$, then 
$P(z)$ is given in (\ref{s11}) for $-1 \le z \le 1$.

Generically, the product distribution can be calculated via the introduction of the Mellin integral transform of the function $f(w)$ for $w \ge 0$,
\begin{equation}
  M\{f(w) |s \} \equiv \int_0^{\infty} w^{s-1} f(w) d w
  \label{MellinTransf}
\end{equation}
where, for real $s$, the Mellin transform $M\{f(w) |s \}$ is simply the $(s-1)$-moment $\left< w^{s-1} \right>$ of the distribution function $f(w)$.
As an example, let us consider the probability distribution of the product of two random variables  each with an independent distribution. Suppose each of the two semi-positively defined random variables $x_1, x_2$ obeys an independent distribution, satisfying $\int_0^\infty f_1 (x_1)\, dx_1 = 1, \ \int_0^\infty f_2 (x_2) \, dx_2 = 1$
\footnote{Although we focus on just semi-positively defined random variables (that smoothly includes zero) for simplicity, the argument can easily be applied to the entire region.}.
Then the product distribution of these two random variables becomes \cite{Huntington:1939}
\begin{equation}
 \begin{split}
  P(z) =& \int_0^\infty \int_0^\infty f_1(x_1)\, f_2 (x_2)\, \delta(x_1 x_2 - z)\, dx_1 dx_2\\
  =& \int_0^\infty {1\over x_2}\, f_1 \left(z \over x_2 \right)\, f_2 (x_2)\, dx_2.
 \end{split}
\end{equation}
When we multiply $z^{s-1}$ and integrate over $z$, we obtain
\cite{Epstein:1948}:
\begin{equation}
 \begin{split}
  M\{ P(z)|s\} = M\{ f_1(x_1)|s\} \cdot M\{ f_2(x_2)|s\}.
 \end{split}
\end{equation}
Therefore the product distribution can be written in term of inverse Mellin integral transforms by
\begin{equation}
 \begin{split}
  P(z) =& {1\over 2\pi i} \int_L z^{-s}  M\{ P(z)|s\} \, ds
 \end{split}
 \label{inverse Mellin transform}
\end{equation}
where the line integral is performed in the complex $s$ plane for $z \geq 0$. This easily generalizes to the case of the product of $n$ semi-positive random variables $x_j$ each with its own distribution $f_j(x_j)$, $z=x_1 \cdots x_n$, where
\begin{equation}
  M\{ P(z)|s\} = \prod_{j=1}^n M\{ f_j(x_j)|s\} 
\end{equation}
Following this and (\ref{MellinTransf}), we see that the expectation value $\left< z^p \right>= M\{ P(z)|p+1\}$ is given by
\begin{equation}
\left< z^p \right>=\prod_{j=1}^n \left< x_j^p \right>
\label{pMoments}
\end{equation}

Using the Mellin integral transformation, we can find the product distribution for normal distribution, $P_{\rm normal}(x_i)=f_i (x_i >0) = \sqrt{(2 /\pi\sigma_i^2)}e^{-x_i^2/2\sigma_i^2}$ \cite{Springer:1966,Springer:1970}.
The resulting distribution goes as
\begin{equation}
 \begin{split}
  P(z) =& N_0^{-1} \,  G^{
  \begin{smallmatrix}{}
   n&0 \\ 0&n
  \end{smallmatrix}
  }
  \left( \left. z^2 \prod_{i=1}^{n} {1\over 2 \sigma_i^2}\right| \underbrace{0,0,\cdots, 0}_{n} \right),\\
  N_0 =& (2\pi)^{n/2} \prod_{i=1}^n \sigma_i,
 \end{split}
 \label{product distribution for normal}
\end{equation}
where the Meijer G-function obeys the following differential equation:
\begin{equation}
 \left[(-1)^{p-m-n} w \prod_{j=1}^p \left(w {d \over dw} - a_j + 1\right) - \prod_{j=1}^q \left(w {d\over dw} - b_j\right) \right]
  G^{
  \begin{smallmatrix}{}
   m&n \\ p&q
  \end{smallmatrix}
  }
 \left( w \left|
   \begin{array}{cccc}
    a_1,&\cdots, &a_p \\
    b_1,&\cdots, &b_q
   \end{array}
  \right. \right)
 =0,
\end{equation}
which can also be expressed in the integral form: 
\begin{equation}
 G^{
  \begin{smallmatrix}{}
   m&n \\ p&q
  \end{smallmatrix}
  }
 \left( w \left|
   \begin{array}{cccc}
    a_1,&\cdots, &a_p \\
    b_1,&\cdots, &b_q
   \end{array}
  \right. \right)
 = {1\over 2\pi i} \int_L w^{-s} {\prod_{j=1}^m \Gamma (s+b_j) \cdot \prod_{j=1}^n \Gamma (1-a_j-s) \over \prod_{j=n+1}^p \Gamma (s+a_j) \cdot \prod_{j=m+1}^q \Gamma (1-b_j-s)} ds,
\end{equation}
where the line integral is performed in the complex plane.
Note that the integration of the Meijer G-function is an inverse Mellin transform.
Then, the set of Meijer G-function relevant for us becomes
\begin{equation}
 G^{
  \begin{smallmatrix}{}
   n&0 \\ 0&n
  \end{smallmatrix}
  }
  \left( w \left| \underbrace{0,0,\cdots, 0}_{n} \right. \right)
  = {1\over 2\pi i} \int_L w^{-s} {\prod_{j=1}^n \Gamma (s)} ds.
  \label{MeijerGn}
\end{equation}
When $n=2$, this function is simply the modified Bessel function of the second kind $K_0 (\sqrt{w})= K_0({z}/{\prod_{j=1}^n}\sqrt{2}\sigma_j)$. The cases for $n=1, 2, 3$ are shown in Figure \ref{fig:product distribution under uniform}.

Finally, the distribution for a product of the same random variable itself ($z = x^n$) can be obtained by equating $\int_0^\infty P(z) dz = \int_0^\infty f(x) dx$ with a distribution $f(x)$ for the random variable $x$.
The resulting power distribution becomes
\begin{equation}
 \begin{split}
  P(z) = {z^{-1 + 1/n} \over n} f(z^{1/n}).
 \end{split}
 \label{product distribution for same parameter}
\end{equation}
Note that a special case with $n=2$ is known as a part of $\chi^2$-distribution, $\chi^2 (1)$.
It is also worth noting that the formula applies even for a non-integral $n$.

  \subsubsection{Asymptotic behavior around $z=0$ \label{sec:asympt-behav-around}}
  
For both the uniform and the normal distributions, we show that the asymptotic behaviors of the product distributions have the same divergent behaviors at $z=0$.
  Since the asymptotic behavior of the product distribution (\ref{product distribution for uniform}) for uniform distribution is explicit, let us elaborate on the asymptotic behaviors of the more non-trivial situations.

  The asymptotic behavior of product distribution (\ref{product distribution for normal}) for normal distribution can be obtained by expanding the Meijer G-function around $z=0$:
  \begin{equation}
   \begin{split}
    P(z) = N_0^{-1} \,  G^{
  \begin{smallmatrix}{}
   n&0 \\ 0&n
  \end{smallmatrix}
  }
  \left( \left. z^2 \prod_{i=1}^{n} {1\over 2 \sigma_i^2}\right| \underbrace{0,0,\cdots, 0}_{n} \right)
    \stackrel{z\sim 0}{\sim}
    {1\over 2 (n-1)! \prod_{i=1}^n \sigma_i} \left({2\over \pi}\right)^{n/2} \left(\ln \frac{1}{|z|} \right)^{n-1} + \cdots.
   \end{split}
  \end{equation}
  Therefore the product of independent random variables diverges as powers of log functions even though the original individual distributions are smooth at zero. Up to the normalization factor, the divergent behavior at $z=0$ is the same as that for the uniform distributions (\ref{product distribution for normal}). This property is intuitively clear. So it follows that if $P(y_1) \sim [\ln (1/y_1)]^{n_1}$ and $P(y_2) \sim [\ln (1/y_2)]^{n_2}$, then the probability distribution of $z=y_1y_2 \ge 0$ goes like, as $z\rightarrow 0$, 
  $$ P(z) \sim [\ln (1/z)]^{n_1 + n_2 +1}$$

However, the diverging behavior changes once some random variables are raised to some powers.
  As we see immediately from (\ref{product distribution for same parameter}), the diverging behavior 
  appears as negative powers of $z$ if the original distribution is smoothly finite at $z=0$.
 To be specific, let us find out what happens in the case of the product $z = x_1^m x_2^n$ where both $x_1$ and $x_2$ have the same uniform distribution from 0 to 1, as an example.
  The distributions for the power of each parameter are given in (\ref{product distribution for same parameter}) with $f(x_i) =1$.
  The resulting distribution can be calculated by
  \begin{equation}
   \begin{split}
    P(z) =& \int_0^1 \int_0^1 {w_1^{-1+1/m} \over m} {w_2^{-1+1/n} \over n} \delta(w_1 w_2 -z)\, dw_1 dw_2\\
    =& {1\over m-n} \left( z^{-1+1/m} - z^{-1+1/n} \right).
   \end{split}
  \end{equation}
  This formula also applies at $m=n$.
  At this limit, the function goes like $z^{-1+1/n} \ln (1/z)$ as $z \rightarrow 0$, which is slightly more diverging than that for just a simple power of a single random variable.
  Repeating this calculation, we can find the probability distribution for $z = x_1^{n_1} \cdots x_m^{n_m} $, the $m$-product of arbitral powers $n_i$ of $x_i$, with each $x_i$ obeying the same uniform distribution from 0 to 1,
  \begin{equation}
   \begin{split}
    P(z) = \sum_{i=1}^{m} {n_i^{m-2} \over \prod_{j \neq i} (n_i-n_j)} z^{-1 + 1/n_i}.
   \end{split}
  \end{equation}
  When we set $n_i = n$ for all $n_i$, the function goes like $z^{-1+1/n} (\ln z)^{m-1}$.
  We expect that the same divergence is obtained in the case of normal distribution.

  Let us summarize the divergent behaviors of the various product distributions in Table \ref{tab:asymptotics of product distribution}.
  The product distributions are in general divergent at $z=0$. The divergence becomes powers if there exists a product of the same random variable inside $z$, which is more divergent at $z=0$ than the logs.
  It is worth noting that the power behavior must be less divergent than $z^{-1}$, which is necessary for $P(z)$ to be properly normalized.

    \begin{table}[t]
   \begin{center}
    \begin{tabular}{|c||c|}
     \hline
     $z$& Asymptote of $P(z)$ at $z=0$ \\
     \hline \hline
     $x_1 \cdots x_n $ & $(\ln (1/|z|))^{n-1}$ \\ \hline
     $x_1^n$ & $z^{-1+1/n}$ \\ \hline
     $x_1^n \cdots x_m^n $ & $ z^{-1+1/n} (\ln (1/|z|))^{m-1}$ \\ \hline
     $x_1^m x_2^n $ & $(z^{-1+1/m} -z^{-1+1/n})/(m-n)$ \\ \hline
     $x_1 \cdots x_m $ / $y_1 \cdots y_n $ & $(\ln (1/|z|))^{m-1}$ \\ \hline
     $x_1^m/ y_1^n$ & $z^{-1+1/m}$\\ \hline
     $x_1^{n_1} + \cdots + x_m^{n_m}$ & $z^{-1 + 1/n_1 + \cdots 1/n_m}$\\ \hline
     $x_1 x_2$, $0< c = x_1/ x_2 < \infty $ & smooth \\ \hline
     $x_1 x_2$, $0\leq c = x_1/ x_2$ or $c\leq \infty$ & $\ln (1/|z|)$ \\ \hline 
    \end{tabular}
   \end{center}
   \caption{\footnotesize The asymptotic behaviors of the probability distributions $P(z)$ of the product, ratio and sum of random variables $x_j$ and $y_j$ which smoothly include zero.
     The product distributions have divergent behaviors at $z=0$ in general.
     The divergent behavior of ratio distribution is simply that given by that of the numerator.   
     The sum distribution works toward erasing the divergent behavior of each term.
     If there is a constraint among some random variables, the divergent behavior of product distribution may be erased. }
   \label{tab:asymptotics of product distribution}
  \end{table}

     \subsubsection{Divergent behavior versus peaking behavior}

\begin{figure}
    \begin{center}
     \includegraphics[width=19em]{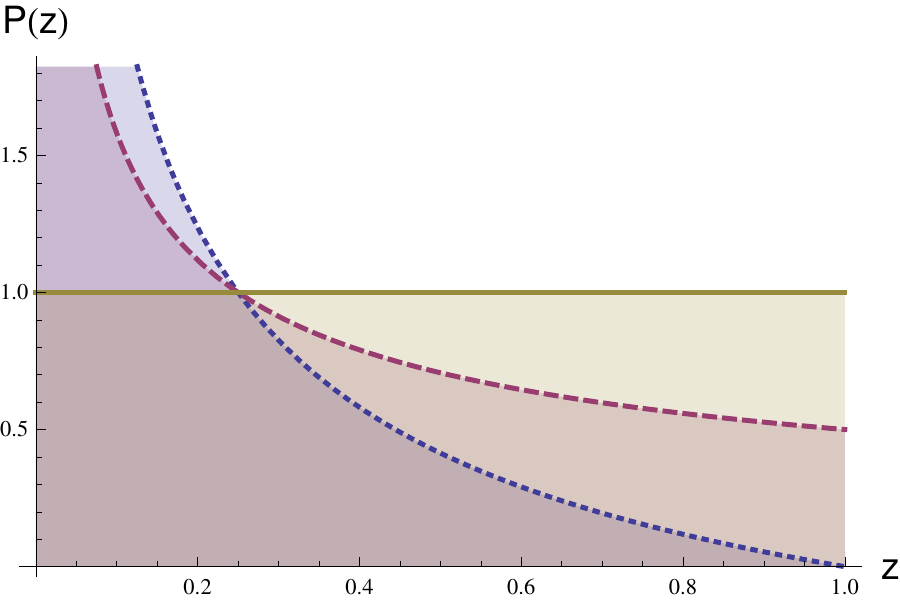}
     \hspace{1em}
     \includegraphics[width=19em]{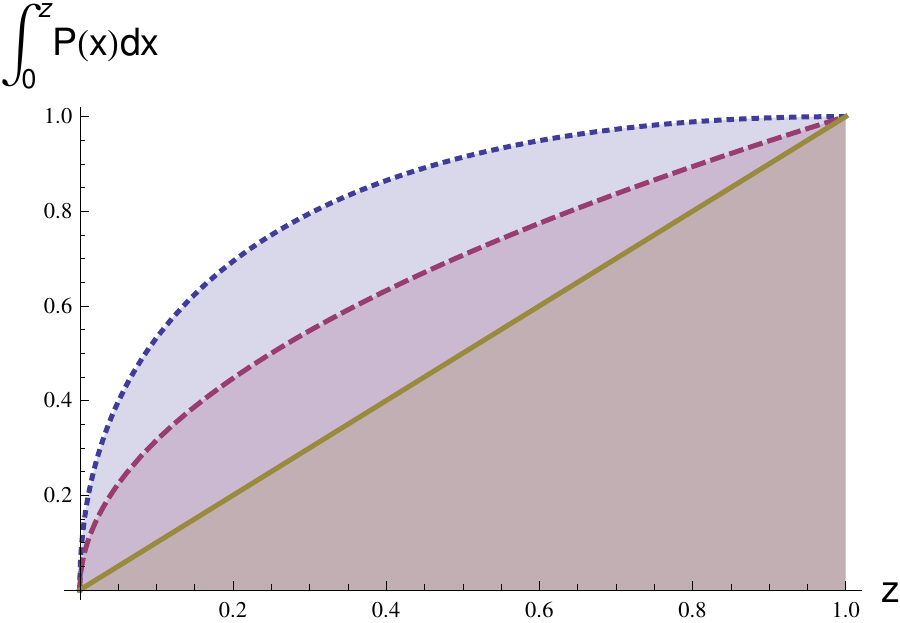}
    \end{center}
    \caption{\footnotesize A comparison of the peakiness (or peaking behavior). The left hand side shows the product distributions for $z$ while the right hand side shows cumulative probability distribution beginning from $z=0$. The curves are for 
    $z = x_1$ (solid horizontal line), $z = x_1^2$ (red dashed curve), and $z=x_1^2 x_2$ (blue dotted curve) in each plot, assuming uniform distributions for $x_1$ and $x_2$.
    .}
    \label{fig:peaking-comparison}
   \end{figure}

In general, a distribution is more peaked at $z=0$  if it is more divergent there, so the divergent behavior at $z=0$ provides a very useful measure of the peakiness (or the peaking behavior) of the distribution. However, they are not equivalent. The divergent behavior is mathematically defined independent of the details while the peakiness does depend on the actual distributions of the random variables. In general, the peakiness becomes stronger for more powers of the random variables.  Instead of providing a definition for the peakiness (we shall introduce a measure on the peakiness suitable for the $\Lambda$ problem at hand in Section \ref{meaL}),  let us illustrate with an example.  For simplicity, let us compare the distribution of $z=x_1^2$ and $z=x_1^2x_2$.
   Assuming uniform distribution from 0 to 1 for $x_1$ and $x_2$, the product distributions are obtained by ($0 \le z \le 1$)
   \begin{equation}
    \begin{split}
     P(z=x_1^2) = {z^{-1/2} \over 2}, \quad P(z = x_1^2 x_2) = z^{-1/2} - 1.
    \end{split}
   \end{equation}
   Although the distributions for both $z=x_1^2$ and $z=x_1^2x_2$ have the same $z^{-1/2}$ divergent behavior at $z=0$, the distribution for $z=x_1^2x_2$  is more peaked there, as we see in Figure \ref{fig:peaking-comparison}. We see that the peakiness depends on both the normalization of the divergent term as well as on the tail behavior (values away from $z=0$).

  As another example,  
  let us calculate the expectation values of the following two cases, assuming uniform distributions for all $x_i \ge 0$,
   \begin{equation}
    \begin{split}
     \left< \prod_{i=1}^{n} x_i \right> =& \int_{0}^{1} dz \, {z \over (n-1)!} \left(\ln {1\over z}\right)^{n-1} = \frac{1}{2^{n}},\\
     \left< x_1^n \right> =& \int_{0}^{1} dz \, z {z^{-1+1/n} \over n} = {1\over n+1}
    \end{split}
    \label{multiple vs. product in vev}
   \end{equation}
   where the first line follows from (\ref{pMoments}).
   Although the product distribution has only a log divergence versus the $z^{-1+1/n}$ power divergence of the power distribution, we see that $P(z)$ for $z=\prod_{i=1}^{n} x_i$ is more peaked than that for $z=x^n$.
    In general, we expect more population in smaller $z$ if $z$ involves more random variables rather than more powers of the same random variable.

    If each uniform distribution has a lower bound $\epsilon$ such that $\epsilon \leq x_i \leq 1$, the expectation value becomes (using (\ref{pMoments})),
    \begin{equation}
     \begin{split}
      \left< \prod_{i=1}^{n} x_i \right> =&  \frac{1}{2^{n}}(1+\epsilon)^n.
     \end{split}
    \end{equation}
    Therefore, if $\epsilon$ is small enough, the expectation value, which may be treated as a measure of the peakiness, is roughly the same as that given in (\ref{multiple vs. product in vev}); however, the probability distribution $P(z)$ no longer diverges. Instead, $P(z) \sim [\ln (1/z)]^{n-1}$ peaks at $z \sim \epsilon$ and then drops to zero at $z=\epsilon^n$.

  \subsection{Ratio distribution}
  
  Next, we describe what happens if random variables come in denominator of $z$.
  First example we consider here is a combined variable given by $z = x_1^m / y_1^n$.
  We consider the parameters $x_1, y_1$ obeying uniform distribution in range from 0 to 1.
  The distributions of the numerator $w_1 = x_1^m$ and the denominator $w_2 = y_1^n$ can be calculated independently following (\ref{product distribution for same parameter}).
  Therefore, the distribution of $z$ is given by
  \begin{equation}
   \begin{split}
    P(z) =& \int_0^1 \int_{0}^{1} P(w_1 = x_1^m)\, P(w_2 = y_1^n)\, \delta \left( {w_1 \over w_2} - z\right) \, dw_1 dw_2\\
    =& \left\{
    \begin{array}[]{l}
     {1\over m+n} z^{-1 + 1/m}  \ {\rm for }\ 0\leq z \leq 1, \\
     {1\over m+n} z^{-1 - 1/n}  \ {\rm for }\ 1\leq z. 
    \end{array}
    \right.
   \end{split}
  \end{equation}
  The divergent behavior at $z=0$ is controlled by the divergence of the numerator, as is intuitively clear.

  Next we consider the other combination given by $z = {x_1 \cdots x_m \over y_1 \cdots y_n}$ where all variables obey uniform distribution in range from 0 to 1.
  The distribution of each numerator $w_1 = x_1 \cdots x_m$ and denominator $w_2 = y_1 \cdots y_n$ is given in (\ref{product distribution for uniform}).
  Again integrating over the distributions with the ratio constraint, we get
  \begin{equation}
   \begin{split}
    P(z) =& \int_0^1 \int_{0}^{1} P(w_1 = x_1^m)\, P(w_2 = y_1^n)\, \delta \left( {w_1 \over w_2} - z\right) \, dw_1 dw_2\\
    =& {(-1)^{-1 + m + n} \over \Gamma(m)} \left(\ln z \right)^{-1 + m + n} U (n, m+n, -2 \ln z) \ {\rm for }\ 0\leq z \leq 1\\
    \stackrel{z \sim 0}{\sim} &  { (-1)^{-1 + m} \over (m-1)! (n-1)! 2^n } \left( \ln z \right)^{m-1} + \cdots,
   \end{split}
  \end{equation}
  where $U(a,b,z)$ is the Tricomi confluent hypergeometric function, defined by
  \begin{equation}
   U(a, b, z) = {1\over \Gamma (a)} \int_0^{\infty} e^{-z t} t^{a-1} (1+t)^{b-a-1}\, dt.
  \end{equation}
  We see again that the divergent behavior is nothing but the one from the numerator.
  Note that we have omitted the expression for $1 \leq z$ because the integration is difficult to be achieved analytically.

 Therefore we may conclude from these examples that the essential divergent behavior in the ratio distribution is dictated by that of the numerator. If the denominator itself has a divergent behavior, that tends to impact on $z$ away from zero. 
   The asymptotic behavior of the ratio distribution is summarized in Table \ref{tab:asymptotics of product distribution}.

  Since the normal distribution has a finite non-zero value at $x_i=0$, just like the uniform distribution case, the essential argument for the divergent behavior should be same.
  So far we have concentrated on just the uniform distributions for each random variable.
  This is because the expression for normal distribution is somewhat more complicated and cannot be achieved analytically in most cases.
  However, if we limit ourselves to particular cases, we can perform the integration even for the normal distribution cases.
  Here we show an example in the case for $z = {x_1 \cdots x_n \over y_1 \cdots y_n}$, where $x_i, y_i$ obey normal distribution centered around 0 with variance 1.
  Let us calculate the simple ratio distribution of $w_1 = x_1 / y_1$ first.
  This is obtained by
 \begin{equation}
  \begin{split}
   P(w_1) = \int_{0}^{\infty} \int_{0}^{\infty} P_{\rm normal} (x_1) P_{\rm normal} (y_1) \, \delta \left({x_1 \over y_1} - w_1 \right) = {2\over \pi} {1 \over 1+ w_1^2}.
  \end{split}
 \end{equation}
 Using this distribution, we can find out the distribution for $z = {x_1 \cdots x_n \over y_1 \cdots y_n}$ as a product distribution iteratively, by
 \begin{equation}
  \begin{split}
   P(z) =& \int \cdots \int P(w_1) \cdots P(w_n) \, \delta \left( w_1 \cdots w_n - z \right)\\
   =& \left\{
   \begin{array}{l}
    {2 (2\ln z) \over \pi^{2k} (-1+z^2)} {1\over (2k-1)!} \prod_{i=1}^{k-1} \left[ (2k-2)^2 \pi^2 + (2 \ln z)^2 \right]\ {\rm for}\ n=2k,\\
    {2\over \pi^{2k-1} (1+z^2)} {1\over (2k-2)!} \prod_{i=1}^{k-1} \left[ (2k-3)^2 \pi^2 + (2\ln z)^2 \right]\ {\rm for}\ n=2k-1.    
   \end{array}
   \right.\\
   \stackrel{z\sim 0}{\sim}& {(-1)^{n-1} 2^n \over \pi^n (n-1)!} (\ln z)^{n-1} + \cdots.
  \end{split}
 \end{equation}
 Therefore we see again that the divergent behavior is nothing but that given in the numerator.

 \subsection{Constrained product distribution}

 If we have a constraint in the product distribution, the divergent behavior will be modified. Properties of a constrained system will come into play in the actual string example to be discussed.
 Since a general analysis is quite complicated, we shall just illustrate some features of the constrained system with a simple example.

 Here we assume $x_1, x_2$ are random variables obeying uniform distribution from 0 to 1.
 Then consider a product distribution of $z = x_1 x_2$, under a constraint $c = x_1/x_2$.
 We defined $c$ as a constant, but which can have a range.
 Let us first calculate the probability distribution function at a particular $(z, c)$ first, 
 \begin{equation}
  \begin{split}
   P(z,c) =& \int_{0}^{1} dx_1 \int_0^{1} dx_2\, \delta (x_1 x_2 - z)\,  \delta \left( {x_1 \over x_2} - c\right)\\
   =& \int^{1/c}_0 dx_2 \, {x_2 \over 2 \sqrt{c z}} \, \delta \left( x_2 - \sqrt{z \over c} \right)\\
   =& {1\over 2c} \quad {\rm for}\  z \leq c \leq 1, \ {\rm or} \ z\leq  {1\over c} \leq  1.
  \end{split}
 \end{equation}

 If $c$ has a range for the constraint, the distribution for $z$ is given after integrating over the region for $c$ with suitable normalization.
 First we assume $1/2 \leq  c \leq 1$, for instance.
 Then the probability distribution function with respect to $z$ becomes
 \begin{equation}
  P \left(z, {1\over 2} \leq  c \leq  1 \right)
   = \left\{
      \begin{array}{l}
       2 \ln 2 \quad {\rm for} \ 0\leq z< {1\over 2},  \\
       2 \ln {1\over z}  \quad {\rm for} \ {1\over 2} < z < 1.
      \end{array}
     \right.
 \end{equation}
 Suppose the upper bound for $c$ is bigger than unity, let us take $1/2 \leq  c \leq 10$ as an illustration; then the probability distribution function is given by
 \begin{equation}
  P \left(z, {1\over 2} \leq  c \leq  10 \right)
   = \left\{
      \begin{array}{l}
       {5 \over 7} \ln 20 \quad {\rm for} \ 0\leq z \leq  {1\over 10},  \\
       {5 \over 7} \ln {2\over z}  \quad {\rm for} \ {1\over 10} < z \leq {1\over 2},\\
       {10 \over 7} \ln {1\over z} \quad {\rm for} \ {1\over 2} < z \leq 1.
      \end{array}
     \right.
 \end{equation}
 As we see from above, generically the distribution does not diverge at $z=0$, if $c$ passes through neither zero nor infinity.
 Although the product distribution suggests a divergent behavior as discussed in Section \ref{sec:asympt-behav-around}, a constraint may work to erase the divergence in the product distribution.

 However if $c$ passes through either zero or infinity, the divergence behavior is retained.
 When ${1/2} \leq c \, (\leq \infty)$, the distribution goes as
 \begin{equation}
  P \left(z, {1\over 2} \leq c\right)
   = \left\{
      \begin{array}{l}
       {2\over 3} \ln {2\over z} \quad {\rm for}\ 0\leq z \leq {1\over 2},\\
       {4\over 3} \ln {1\over z} \quad {\rm for}\ {1\over 2} < z \leq 1.
      \end{array}
     \right.
 \end{equation}
 While, if $0\leq c \leq 1/2$, we get
 \begin{equation}
  P \left(z, 0 \leq c \leq {1\over 2}\right) = 2 \ln {1\over z} \quad {\rm for}\ 0\leq z \leq {1\over 2}.
 \end{equation}
 We see that the probability distribution function diverges as $\ln z$ when $c$ passes through either zero or infinity, or both.
 This generic feature of the constrained system is summarized in Table \ref{tab:asymptotics of product distribution}.

  \subsection{Sum distribution} 
  
  So far we have explained what happens if the distribution is given as a product of some random parameters.
  Here, we shall elaborate on the sum distribution.
  This sum distribution is crucial if the population of vacuum consists of summation of a number of quantities, e.g. quantized fluxes considered in \cite{Bousso:2000xa}.

  The sum distribution is understood rather easier than the product distribution.
  If there are two independent random parameters satisfying $\int_0^\infty f_1 (x_1)\, dx_1 = 1, \ \int_0^\infty f_2 (x_2) \, dx_2 = 1$ (finite range for uniform distribution), the sum distribution of $z = x_1 + x_2$ becomes
  \begin{equation}
   \begin{split}
    P(z) =& \int_0^\infty \int_0^\infty f_1(x_1)\, f_2 (x_2)\, \delta(x_1 + x_2 - z)\, dx_1 dx_2\\
    =& \int_0^\infty f_1 (z-x_2) \, f_2(x_2) \, dx_2.
   \end{split}
  \end{equation}

  Let us apply this formula to a case which consists of a summation of the arbitrary powers of uniformly distributed random variables, where each term has the divergent behavior given by the distribution (\ref{product distribution for same parameter}).
  The probability distribution of the $m$-summation of $n_i$-powers of the random variables, $z=x_1^{n_1} + \cdots + x_m^{n_m}$ is given by
  \begin{equation}
   \begin{split}
    P(z) =& \int_0^1 \prod_{i=1}^m {w_i^{-1+1/n_i} \over n_i} \, \delta (w_1 + \cdots + w_m - z)\, dw_1 \cdots dw_m\\
    =& {\prod_{i=1}^m \Gamma(1+1/n_i) \over \Gamma (\sum_{i=1}^{m} 1/n_i)} z^{-1 + \sum_{i=1}^m 1/n_i}.
   \end{split}
  \end{equation}
  Thus we see that the singular behavior at $x_j=0$ for individual terms tend to be smoothed out at $z=0$ as the number of terms $m$ in the summation increases.
  For instance, if $n_1 = n_2 = 2$, then summation of two terms are enough to suppress the divergent behavior.
  This smoothing behavior in the sum of product distributions also applies to the other distributions.
  The combined distributions including the sum distribution is summarized in Table \ref{tab:asymptotics of product distribution}.

\section{The Bousso-Polchinski example\label{sec:bousso-polch-example}}

  Let us move on to the population argument in the string landscape based on the flux quantization \cite{Bousso:2000xa} (see also a review \cite{Douglas:2006es,Bousso:2012dk}). In this toy model,
  the cosmological constant is given as a combination of bare constant $\Lambda_{\rm bare}$ and four-form fluxes contribution, by
  \begin{equation}
  \label{BPsum}
   \begin{split}
    \Lambda = \Lambda_{\rm bare} + {1\over 2} \sum_{i=1}^{J} n_i^2 q_i^2
   \end{split}
  \end{equation}
  where $q_i$ are basic quantity assigned for each cycle after compactification, and $n_i$ are quantized integers related to fluxes of the cycles.
  A simple version of this model was constructed in \cite{Brown1988,Brown1987}.
  
\begin{figure}[t]
   \begin{center}
    \includegraphics[width=20em]{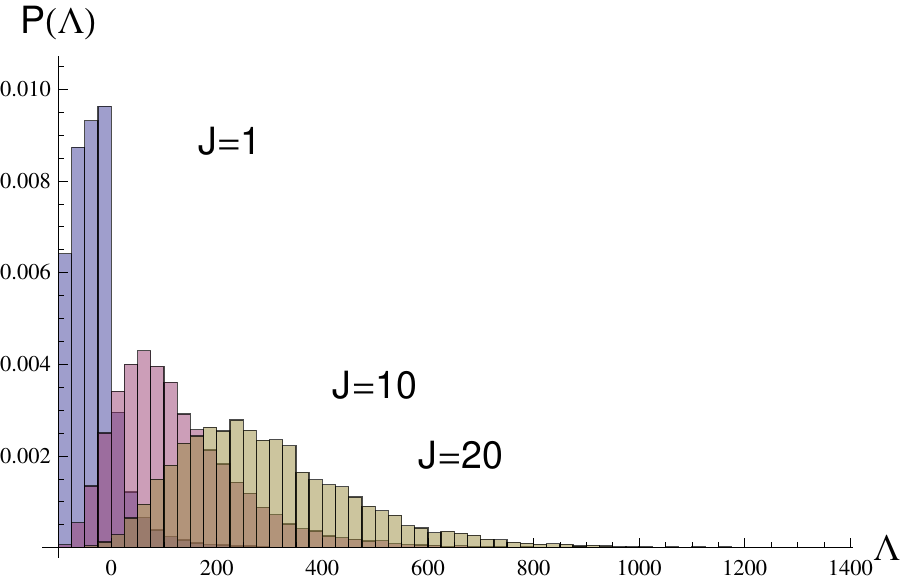}
   \end{center}
   \caption{\footnotesize The probability distribution of $\Lambda$ in the Bousso-Polchinski type (\ref{BPsum}).
   The probability distribution in the $J=1$ ($J=10, 20$) case peaks at zero ($\sim 50, 200$). 
   Population moves toward large $\Lambda$ as the number of types of fluxes $J$ increases.
   As a result, tiny cosmological constant is allowed but not preferred in the presence of more than one cycle.}
   \label{fig:Bousso-Polchinski}
  \end{figure}

  First, for fixed $q_j \ne 0$ and $\Lambda_{\rm bare}=0$,  we can read off from Table \ref{tab:asymptotics of product distribution} that the resulting distribution does not have a peaking preference for $\Lambda=0$. Adding a negative 
  $\Lambda_{\rm bare}$ will not introduce a peaking behavior at $\Lambda =0$. 
  It should be intuitively obvious that the situation would not improve in general. To see this more explicitly, let us impose randomness for the parameters $\Lambda_{\rm bare}, n_i, q_i$ and see how it affects in the probability distribution of $\Lambda$. To be specific, 
  we assume that the $q_i$ obey normal distributions centered around 0 with variance 1, and $\Lambda_{\rm bare}$ obeying uniform distribution from $-100$ to 0.
  Since the $n_i$ are quantized numbers, we give random integer values for $n_i$ from 0 to 10.
  Using these distributions, we plot the histograms of $\Lambda$ at $J=1, 10, 20$ respectively in Figure \ref{fig:Bousso-Polchinski}.

 The histogram illustrates our argument for the sum distribution: although the distributions of quadratic power itself for individual terms give us diverging behavior at $\Lambda = 0$ point (notice that $\Lambda = 0$ is not the origin of the histogram in Figure \ref{fig:Bousso-Polchinski}).
  Once there are more than one term in the sum in (\ref{BPsum}), the divergent behavior is being smoothed out.
  The preference of large cosmological constant is also an outcome of sum distribution, because all terms related fluxes are positively defined.
  But if we focus on $J=1$ case, $\Lambda = 0$ is mostly preferred among the population for $\Lambda >0$ owing to the divergent behavior of the product distribution for $z=n_1^2 q_1^2$.
  However, as we expect a number of non-trivial cycles in Calabi-Yau compactifications easily, such the divergent behavior is quite likely smoothed out, and no preference for $\Lambda = 0$ remains in this type of models.
 A similar plot was obtained for the $J=7$ case in \cite{SchwartzPerlov:2006hi}.

More generally, suppose the effective potential involves 2 or more components so that each involves only a subset of the moduli, e.g.,
  \begin{equation}
V(\phi_i, \varphi_j) = V_1(\phi_i) + V _2(\varphi_j)
  \end{equation}
 where the two sets of moduli do not couple to each other.
Even if the minimum of $V_1$ is peaked at $V_1=0$ and the minimum of $V_2$ is peaked at $V_2=0$, it is possible that there is no peaking at the minimum at $V=0$. So, for $\Lambda=0$ to be preferred, the sets of moduli should be coupled together. In general, if we include the interaction (loop contributions and $\alpha '$ corrections), all moduli should couple (at least indirectly) to each other.

  Since moduli stabilization requires some relations in all terms in the potential for both extremal conditions and stability constraints, we will argue how product distribution and its divergence behavior dominate the population in the stringy landscape while freezing randomness for summations, especially using a class of model in type IIB.
It is possible that more coupled moduli in $V$ will lead to a stronger peak at $\Lambda=0$. This feature obviously depends on the detailed dynamics.

\section{The K\"ahler uplifting in the large volume scenario in Type IIB theory\label{sec:large-volume-scen}}

In this section, we like to study a specific example in string theory, in particular how the extremum and stability constraints impact on the basic idea of preference for small $\Lambda$s.
The formulae obtained in this section will be used for the probability distribution of $\Lambda$ in Section \ref{meaL} .
Since supersymmetry can remain for negative $\Lambda$, their presence may mask the feature we are interested. So let us focus on the positive $\Lambda$ case. and then comment on the negative $\Lambda$ case. As we shall see, the case we study automatically extends to negative $\Lambda$s.

There are some efforts to understand the potential for moduli stabilization in type II string theory.
For necessity condition toward stabilization of all moduli, we may focus on two universal moduli consisting of a K\"ahler and dilaton.
The extremum \cite{Hertzberg:2007wc,Wrase:2010ew} at positive $\Lambda$ and stability \cite{Shiu:2011zt} constraints are studied to limit models in string theory (see also an application for higher dimensional $dS$ vacua \cite{VanRiet:2011yc}).

To find the probability distribution of $\Lambda$, we have to start with an explicit relatively simple model that has a de-Sitter vacuum solution. Here we choose to illustrate the peaking behavior with a single modulus model with K\"ahler uplifting in the large volume scenario (LVS) in IIB string theory \cite{Balasubramanian:2004uy,Westphal:2006tn,Rummel:2011cd,deAlwis:2011dp,Balasubramanian:2005zx}, focusing on positive $\Lambda$ vacuum solution.
We show how the peaking behavior of a product distribution appears when the potential has more than one term in it. We show also that the constraints typically modifies but not erase the peaking behavior.  Let us briefly review the large volume scenario before focusing on the single modulus case.

One may consider the KKLT scenario \cite{Kachru:2003aw} instead of LVS.
KKLT introduces a non-perturbative term in the superpotential to stabilize at supersymmetric vacuum, therefore we may expect the tree level contribution and the non-perturbative contribution to be comparable: $W_0 \sim W_{\rm np}$.
 Thus $W_0$ is expected to be exponentially small.
The stable positive $\Lambda$ vacuum is achieved by adding some $\overline{\rm D3}$-branes at the tip of throat as an uplifting term
\footnote{The backreaction of $\overline{\rm D3}$-branes has been recently discussed in \cite{DeWolfe:2008zy,McGuirk:2009xx,Bena:2009xk,Bena:2010ze,Dymarsky:2011pm,Bena:2011hz,Bena:2011wh,Massai:2012jn}. See also the argument of $\overline{\rm D6}$-branes \cite{Blaback:2011nz,Blaback:2011pn,Blaback:2012nf}.}.
On the other hand, LVS allows the relation $W_0 \gg W_{\rm np}$ so that we can consider naturally large $W_0$ after stabilizing the complex moduli.
This relation gives us an advantage to make the potential relatively simple, especially at smaller $\Lambda$.

\subsection{The K\"ahler uplifting in the large volume scenario}

The large volume scenario is present in type IIB string theory.
Owing to the {\it no-scale structure}, additional effects beyond the classical potentials are helpful to stabilize the K\"ahler moduli, while complex structure moduli can be stabilized at tree level.
Here we consider an additional non-perturbative effect in the superpotential, originally introduced in \cite{Kachru:2003aw}, and also $\alpha'$-correction in the K\"ahler potential \cite{Becker:2002nn}.
These two effects break the no-scale structure at next leading order in the potential, therefore K\"ahler moduli can now be stabilized.
This fact enables us to impose a mass hierarchy between the K\"ahler moduli and the complex structure moduli so that we can safely assume that complex moduli are stabilized at higher scales.
Thus we focus only on the the K\"ahler sector in this paper.
More detailed analysis in the presence of stringy loop corrections were considered systematically in \cite{Cicoli:2007xp,Cicoli:2008va,Berg:2007wt}.
Some advantages for phenomenological applications of LVS in terms of mass scale hierarchies are available in \cite{Conlon:2005ki,Blumenhagen:2009gk}.

The model is clarified to have a $dS$ stable vacuum in a particular parameter region of minimal configuration \cite{Rummel:2011cd,Westphal:2006tn,Parameswaran:2007kf}, satisfying the large volume assumption which is important to deal with the $\alpha'$-correction suitably (note that a K\"ahler transformation in the system can increase the volume easily)
\footnote{See also \cite{Cicoli:2012fh} for the argument in the presence of a dilatonic non-perturbative term.}.
A $dS$ minimum is achieved because a term related to the $\alpha'$-correction can contribute positively in the potential.
Such a possibility was also analyzed along the Goldstino direction \cite{Covi:2008ea}.

The effective potential for the K\"ahler moduli is given by
\begin{equation}
 \begin{split}
  V =& e^{K \over M_P^2} \left(K^{I \bar{J}} D_I W D_{\bar{J}} \bar{W} - {3\over M_P^2}|W|^2\right),\\
  K =& -2 M_P^2 \ln \left({\cal V} + {\hat{\xi} \over 2} \right),\\
  \hat{\xi} =& - {\zeta (3) \over 4 \sqrt{2} (2\pi)^3 g_s^{3/2}} \, \chi (M)
  \sim 8.57 \times 10^{-4}\, {\chi (M) \over g_s^{3/2}},
 \end{split}
 \label{LVS potential}
\end{equation}
where $\chi (M)$ is Euler number of the manifold given by $\chi (M) = 2 (h^{1,1} - h^{2,1})$, which can be negative.
Since we assume these complex moduli were already frozen at high scale, we consider $\hat{\xi}$ as just a parameter here though $\hat{\xi}$ has the dilaton dependence.
It is worth noting that the ${\cal V}$ is a dimensionless volume measured in the $\alpha'$ unit.
If we recover the dimensionality of the volume by redefining ${\cal V} = \vol / \alpha'^3$, we see immediately that the $\hat{\xi}$ parameter is  suppressed by $\alpha'^3$ relative to $\vol$, so it may be treated as an $\alpha'$ correction.
Note also that the Planck scale is related to string length by $M_P^2 = {4\pi {\cal V} / (g_s^2 (2\pi)^2 \alpha')}$ where we consider ${\cal V}$ as given after moduli stabilization so that $M_P$ is the constant.
If we fix $\alpha'$ as motivated by the string theory setups, the Planck scale is defined differently depending on the ${\cal V}$; that is, it depends on the vacuum expectation values of the moduli fields.
However, since our focus is the cosmological constant observed in Planck unit, we shall consider the dimensionless ratio $V/M_P^4$, in which the Planck scale dependence is cancelled out.

We consider the volume of this model as that of the {\it Swiss-cheese type}, defined by \cite{Conlon:2005ki,Westphal:2006tn,Rummel:2011cd}
\begin{equation}
 \begin{split}
  {\cal V} =& {1\over 6} \kappa_{iii} (v^i)^3 = \gamma_1 (T_1 + \bar{T_1})^{3/2} - \sum_{i=2} \gamma_i (T_i + \bar{T}_i)^{3/2},\\
  t_i \equiv& {\rm Re} T = \partial_{v^i} {\cal V}, \quad   \gamma_i = {1\over 6 \sqrt{|\kappa_{iii}|}}
 \end{split}
\end{equation}
where the self-intersection number $\kappa_{iii}$ are defined $\kappa_{111} >0$ while $\kappa_{iii} <0 \ (i>1)$.
The superpotential includes non-perturbative terms for the $h^{1,1}$ number of K\"ahler moduli:
\begin{equation}
 \begin{split}
  W =& M_P^3 \left( W_0 + \sum_{i=1}^{h^{1,1}} A_i e^{-a_i T_i} \right).
 \end{split}
\end{equation}
The constants $a_i$ are given depending on the way of getting the non-perturbative terms: $a_i = 6 \pi /\beta_i$ by gaugino condensation on D7-branes, where $\beta_i$ are the one-loop beta function coefficients of the condensing gauge theory ($\beta_i = 3 N_c$ for $SU(N_c)$), and $a_i = 2\pi$ by an instanton effect on Euclidean D3-branes, for instance.
We may take $a_j=2 \pi/N_c$ for integers $N_c > 1$.
The parameters $W_0, A_i$ are functions of complex moduli and open string moduli.
Here we treat them as constant parameters owing to the assumption that the complex moduli are stabilized at higher scale.

\subsection{The single modulus model}

Let us consider the single modulus case of the above scenario, with only the modulus $T_1$.
This case has been studied analytically in \cite{Rummel:2011cd}. It is easy to see that we have a minimum at the origin in the axionic direction of $T_1$, so we shall set ${\rm Im}\, T_1=0$ and consider only the real part $t_1 = {\rm Re}\, T_1$.
Here we set $\alpha' = 1$ for simplicity.
First, let us expand the potential as
\begin{equation}
 \begin{split}
  {V \over M_P^4} \times {{\cal V}^2 \over W_0^2} =& {\cal O} \left( {A_1 e^{-a_1 t_1} \over W_0} \right) + {\cal O} \left( {\hat{\xi} \over {\cal V}} \right)\\
  &+ {\cal O} \left( {A_1 e^{-a_1 t_1} \over W_0} \right)^2  + {\cal O} \left({\hat{\xi} \over {\cal V}} \right)^2  + {\cal O} \left({A_1 e^{-a_1 t_1} \over W_0} \times {\hat{\xi} \over {\cal V}} \right) + \cdots.
 \end{split}
 \label{effectively expanded LVS potential}
\end{equation}
If the potential is controlled by small factors like $A_1 e^{-a_1 t_1}/W_0$ and $\hat{\xi}/{\cal V}$, we can focus on first two terms in the approximate potential \cite{Rummel:2011cd},
\begin{equation}
 \begin{split}
  {V \over M_P^4} \sim& {a_1 A_1 e^{-a_1 t_1} W_0 \over 2 \gamma_1^2 t_1^2} + {3 W_0^2 \hat{\xi} \over 64 \sqrt{2} \gamma_1^3 t^{9/2}}\\
  =& - {W_0 a_1^3 A_1 \over 2 \gamma_1^2} \left( {2 C \over 9 x^{9/2}} - {e^{-x} \over x^2} \right),\\
  C \equiv& {-27 W_0 \hat{\xi} a_1^{3/2} \over 64 \sqrt{2} \gamma_1 A_1}, \quad x \equiv a_1 t_1.
 \end{split}
 \label{Westphal's potential}
\end{equation}
where we redefine several parameters. (Comparing to (\ref{Leg}), we see that $A=-A_1$.) The existence of $dS$ solutions requires
\begin{equation}
-W_0A_1 \ge 0
\end{equation}
which we shall impose together with $\hat \xi >0$. {\footnote {As we shall see, this $-W_0A_1 \ge 0$ condition also allows $AdS$ stable solutions, while $W_0A_1>0$ allows $AdS$ but not $dS$ stable solutions. So allowing $W_0A_1>0$ will not change, up to an overall normalization, $P(\Lambda)$ for $\Lambda \ge 0$, but will change $P(\Lambda)$ for $\Lambda <0$.}}

\subsection{$\Lambda$ for the stable solution}

Now both the extremal condition and the stability constraint can be solved analytically.
The derivatives of the potential are given by
\begin{equation}
 \begin{split}
  \partial_x {V \over M_P^4} =& - {W_0 a_1^3 A_1 \over 2 \gamma_1^2} \left(-{C \over x^{11/2}} + {e^{-x} (x+2) \over x^3}\right),\\
  \partial_x \partial_x {V \over M_P^4} =& - {W_0 a_1^3 A_1 \over 2 \gamma_1^2} \left({11 C \over 2 x^{13/2}} - {e^{-x} (x^2 + 4 x +6) \over x^4}\right). 
 \end{split}
\end{equation}
The $C$ and $x$ are related through the extremal condition $\partial_x V = 0$:
\begin{equation}
 C = x^{5/2} (x+2) e^{-x}.
\end{equation}
Combining with the semi-positive condition $V \ge 0$ and the stability constraint $\partial^2 V >0$, $C$ and $x$ should be within the ranges
\begin{equation}
 3.65 \lesssim C \lesssim 3.89, \quad 2.50 \leq x \lesssim 3.11.
  \label{condition for C}
\end{equation}
where the lower bound is for $V \ge 0$ and the upper bound is for $\partial^2 V >0$

\begin{figure}
 \begin{center}
  \includegraphics[width=20em]{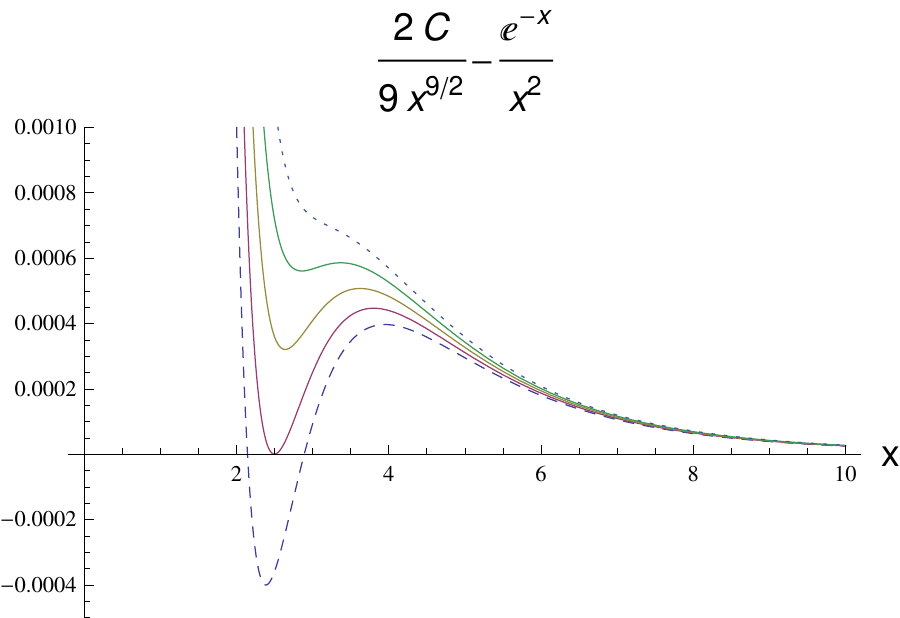}
 \end{center}
 \caption{\footnotesize The plot shows behaviors of the bracket in (\ref{Westphal's potential}) (see \cite{Rummel:2011cd}) at $C\sim 3.55, 3.65, 3.75, 3.85, 3.95$ from bottom to top respectively, for an illustration of the stability constraint.}
 \label{fig:behavior of the bracket}
\end{figure}

The stability of the potential $V(x)$ (up to an overall factor) is shown in Figure \ref{fig:behavior of the bracket}, where we plot the behaviors of the bracket in (\ref{Westphal's potential}).
The  constraints $V \ge 0$ and $\partial^2 V >0$ force the bracket in (\ref{Westphal's potential}) to satisfy
\begin{equation}
 0 \leq \left( {2 C \over 9 x^{9/2}} - {e^{-x} \over x^2} \right) \lesssim 6.25 \times 10^{-4}.
  \label{constraint for the bracket part}
\end{equation}
This bracket in the potential (\ref{Westphal's potential}) can reach negative values if we allow $x< 5/2$.

Next we consider the minimum of the potential (\ref{Westphal's potential}), especially at the vicinity of $\Lambda =V_{\rm min} = 0$, where $C \sim 3.65, \, x= 5/2$.
By using the extremal constraint (\ref{condition for C}), we can expand the minimum of the potential (\ref{Westphal's potential}) around zero cosmological constant:
\begin{equation}
 \begin{split}
 {\Lambda \over M_P^4} \sim& {e^{-5/2} \over 9} \left({2\over 5}\right)^2 {-W_0 a_1^3 A_1 \over \gamma_1^2} \left(x-{5\over 2}\right) + \cdots \\
  =& {1\over 9} \left({2\over 5}\right)^{9/2} {-W_0 a_1^3 A_1 \over \gamma_1^2} \left( C - {225 \sqrt{10} \over 16 e^{5/2}}\right) + \cdots,\\
  C \sim& \left({5\over 2}\right)^{5/2} e^{-5/2} (x+2) + \cdots.
 \end{split}
 \label{Westphal potential at the vicinity of zero}
\end{equation}
For generic values of the parameters $\gamma_1, a_1, A_1, W_0$ and $\hat{\xi}$, $\Lambda/M_P^4$ may be small, say, of order of $10^{-5} - 10^{-4}$. One can always tune $x$ (and $C$) so that  $\Lambda/M_P^4 \sim 10^{-120}$, but this is fine tuning.

Before going to the random variable analysis, we would like to comment on some features.
If we focus on the non-perturbative term from $SU(N_c)$ gaugino condensation on D7-branes, we have $a_1 = 2\pi/N_c$.
 From the stability constraint for de-Sitter vacua (\ref{condition for C}), we may roughly estimate $C \sim 4, \ x \sim 3$.
 Therefore the magnitude of compactified volume is roughly given by,
 \begin{equation}
  t_1 \sim {3\over 2\pi} N_c, \quad {\cal V} \sim \gamma_1 \left(3\over 2\pi \right)^{3/2} N_c^{3/2}.
 \end{equation}
 Thus the volume is easily getting larger as $N_c$ increases.
 Solving $A_1$ (using $C\sim 4$) in the potential (\ref{Westphal potential at the vicinity of zero}), we  obtain
 \begin{equation}
  \begin{split}
   {\Lambda \over M_P^4} \sim {1\over 9} \left({2\over 5}\right)^{9/2} {27 \times 0.1 \over 64 \sqrt{2} \times 4} \left({3\over 2\pi}\right)^{9/2} {W_0^2 \hat{\xi} \over \gamma_1^3} N_c^{-9/2} \sim 4.82 \times 10^{-7} {W_0^2 \hat{\xi} \over \gamma_1^3} N_c^{-9/2}.
  \end{split}
 \end{equation}
 Therefore a large $N_c$ yields a small cosmological constant. However, just taking $N_c$ very large is a fine-tuning in the model.
 Instead of fine-tuning, we randomize the parameters in several ways, and analyze the statistical behavior in this model.

\section{The probability distribution of $\Lambda$}\label{meaL}

The question we are finally ready to address is:
 \begin{quotation}
\noindent  {\it Without fine tuning, what is the preferred values of $\Lambda$, if the parameters in the model are treated as random variables ?}
 \end{quotation}
Here, our discussion is restricted to the single modulus model just reviewed.

\subsection{Probablity distributions of $W_0$ and $A_1$}

For the stringy mechanism to work in this model, it is crucial that the parameters $W_0$ and $A_1$ can take zero values. So let us make some remarks here about zeros of $W_0$ and $A_1$, which depend on the complex structure, the dilaton,  positions of D3-branes, etc.. In general, it is simplest and probably most natural to have $W_0=A_1=0$, yielding supersymmetric Minkowski solutions. So we do expect zeros for them to be very likely. Since other values are also possible, they will have some non-trivial probability distributions, but these distributions should naturally peak at zero values.

Though $W_0$ is different in different models, we focus on a simple example of $T^6$ \cite{Lust:2005dy,Westphal:2006tn,Rummel:2011cd}.
After solving the equations for supersymmetric minima, we get 
\begin{equation}
\label{zeroW0}
W_0 = 2 d u (1-s)
\end{equation}
 where $d u$ is a combination of two parameters involving the expectation values of the real parts $u_i$ of the complex structure moduli $U_i$, and $s$ is the expectation value of the real part of the dilaton \cite{Rummel:2011cd}.
As we see immediately, the $W_0$ can pass through zero easily, as a result of $s=1$ or the combined parameter $d u = 0$.
Even in this simple example, we see that $W_0=0$ is allowed in general. Furthermore, because of the allowed zeros in $du$ and $1-s$, we expect the probability distribution $P(W_0)$ of $W_0$ to be peaked at $W_0=0$.

On the other hand, the dependence of the position of D3-branes in $A_1$ was discussed in \cite{Baumann:2006th}.
The resulting dependence goes like 
\begin{equation}
\label{zeroA1}
A_1 = \hat{A}_1(U_i) \left(f(X_i)\right)^{1/n}
\end{equation}
where $\hat{A}_1(U_i)$ is a function of the complex structure moduli $U_i$ and $f(X_i)$ specifies the positions of D3-branes.
For instance, in the conifold, we get
\begin{equation}
f(X_i) = \prod_{i=1}^{4} X_i^{p_i} - {\mu}^q
\end{equation}
where $p_i \in {\mathbb{Z}}, \, q = \sum p_i$, and $\mu$ specifies the location of the D7-branes in the throat (note that we used the different notation from \cite{Baumann:2006th} to avoid confusions in the other sections).
The case $f(X_i) = 0$ is known as {\it Ganor zero}, which is the special case when 3-branes dependence on the superpotential turns out to be zero \cite{Ganor:1996pe}. So the $A_1$ parameter itself naturally passes through zero in string compactification.
Since we may expect allowed zeros in $\hat{A}_1$ also, the probability distribution $P(A_1)$ of $A_1$ may peak at $A_1=0$. For simplicity, we shall assume in the discussion below that both $W_0$ and $A_1$ have uniform distributions that include zero. This is a rather conservative input.

\subsection{The distribution of $\Lambda$}

Let us define
\begin{equation}
 \begin{split}
  w_1 = {a_1 \over \gamma_1^{2/3}}, \quad w_2 = A_1, \quad w_3 = - W_0, \quad w_4 = \hat{\xi}, \quad
  c = {w_1^{3/2} w_3 w_4 \over w_2} = {64 \sqrt{2} \over 27} C.
 \end{split}
\end{equation}
so $\Lambda$ (\ref{Westphal potential at the vicinity of zero}) can be rewritten in terms of the dimensionless ${\hat \Lambda}$,
\begin{equation}
 \begin{split}
 {\hat \Lambda} \equiv {\Lambda \over M_P^4} \sim { 3\over 2500 \sqrt{5}} {w_1^3 w_2 w_3} \left( c - c_0 \right), \quad
  c_0 = {200 \sqrt{5} \over 3 e^{5/2}} \sim 12.2.
 \end{split}
 \label{lambda with coefficient}
\end{equation}
Neglecting the overall constant factor in ${\hat \Lambda}$, we reduce the problem down to the constrained product distribution given by
\begin{equation}
 {\hat{\Lambda}} \propto w_1^3 w_2 w_3 (c-c_0), \quad c = {w_1^{3/2} w_3 w_4 \over w_2}, \quad c_0 \leq c \leq c_1,
  \label{constrained product distribution from Westphal model}
\end{equation}
where $c_0 \sim 12.2$ as given in (\ref{condition for C}) for the potential (\ref{Westphal's potential}).
Since here we restrict ourselves in the region for smaller cosmological constant, $c_1$ cannot be so large, but we allow $c$ to have some range.
The value of $c_1$ fixes the coefficient of the most divergent term, and is irrelevant for the divergent behavior itself as long as $c_1$ is finite. Stability condition (\ref{condition for C}) requires that $c_1\lesssim 13.0$.

Now we start imposing randomness in the constrained product model (\ref{constrained product distribution from Westphal model}).
Although we can deal with $w_1, w_2, w_3$ as random parameters passing through zero, there is a concern about $w_4 = \hat{\xi}$.
The definition of $\hat{\xi}$ include dilaton dependence as in (\ref{LVS potential}), therefore the parameter may be treated as random parameter in a range of finite values except for zero.
This is because zero-ness of $\hat{\xi}$ is given as a result of zero value of Euler number, therefore it is difficult to expect that $\hat{\xi}$ smoothly passes through zero.
If there is a parameter in a product which does not pass through zero, that parameter is irrelevant for the divergent behavior of the product at $\Lambda=0$.
Thus we neglect the dependence of $w_4$ for the moment. We shall include $w_4$ back in Case 1 in Appendix \ref{sec:other-scen-param}.

We may further consider the possibility that the $a_1$ parameter has a lower cutoff, as a result of the size of the gauge group which contributes to the gaugino condensation on D7-branes.
  If the parameter $a_1=0$ is excluded, $a_1$ (or $w_1$) cannot contribute to the divergent behavior in the constrained product distribution in (\ref{constrained product distribution from Westphal model}).
  Instead of treating $w_1$ as a random parameter, we consider $SU(N_c)$ gaugino condensation on D7-branes to generate the non-perturbative terms with fixed $N_c$ and set $\gamma_1 = 1$ so that $w_1 = 2\pi/N_c$ for simplicity. We shall treat $a_1$ (or $w_1$) as a random parameter (in 2 different ways) in Case 2 and Case 3 in the Appendix \ref{sec:other-scen-param}.

  \begin{table}
   \begin{center}
    \begin{tabular}{|c|c|c|c|}
     \hline
     Model & Random variables & Asymptote & $\sqrt{\left< \hat{\Lambda}^2 \right>}$ \\
     \hline\hline
     (\ref{simple random model from westphal}) & $w_2, w_3$ & $ - \ln \hat{\Lambda}$ & $1.09 \times 10^{-5} \hat{\xi} (2 \pi / N_c)^{9/2}$\\
     \hline
     (\ref{constrained product distribution with xi}) & $w_2, w_3, w_4$ & $ - \ln \hat{\Lambda}$ & $7.72 \times 10^{-6} (2 \pi / N_c)^{9/2}$\\
     \hline
     (\ref{constrained product distribution without xi}) & $w_1, w_2, w_3$ & $\hat{\Lambda}^{-4/9}$ & $5.09 \times 10^{-6} \hat{\xi} $ \\
     \hline
     (\ref{including peaking in a_1}) & $1/y_m \leq w_1=1/y_1 \leq 1, w_2, w_3$ & $- y_m^4 \ln \hat{\Lambda}$ & $2.50 \times 10^{-6} \hat{\xi}$ ($y_m \rightarrow \infty$)\\
     \hline
    \end{tabular}
   \end{center}
   \caption{\footnotesize A summary of the peaking property in the constrained product distribution (\ref{constrained product distribution from Westphal model}) for the 4 plausible scenarios. The second column shows the parameters that are being treated as random variables.  
  We classify the asymptote of probability distribution $P(\Lambda)$ and the square root of the variance $\sqrt{\left< \hat{\Lambda}^2 \right>}$ for each set of random variables among $w_1, w_2, w_3, w_4$.
  They have uniform distribution except for $w_1$ in the last case (\ref{including peaking in a_1}) where $y_1=1/w_1$ has a uniform distribution.}
   \label{tab:asymptote and vev of LVS}
  \end{table}

This means that there are a number of possibilities that we should consider.
We study plausible scenarios and summarize the results in Table \ref{tab:asymptote and vev of LVS}.
Below we discuss the simplest scenario and relegate the other three scenarios to Appendix \ref{sec:other-scen-param}.
We see that each scenario yields a peaking behavior (at zero) for the probability distribution of $\Lambda$, though the specific divergent behavior differs under different assumptions.

\subsection{A simple scenario for the parameters in $V$}
 
    With the above simplifying assumptions, $\hat \Lambda = {3z}/{2500 \sqrt {5}}$ where
  \begin{equation}
   \begin{split}
    z= \left({2\pi \over N_c}\right)^3 w_2 w_3 (c-c_0), \quad c = \hat{\xi} \left( {2\pi \over N_c}\right)^{3/2}  {w_3 \over w_2}, \quad
    c_0 \leq c \leq c_1.
   \end{split}
   \label{simple random model from westphal}
  \end{equation}
 That is, we treat only $w_2=A_1$ and $w_3=-W_0$ as random variables while the other quantities are fixed parameters.  Performing the conditioned integrals, we get
  \begin{equation}
   \begin{split}
    P(z,c) =& \int dw_2 dw_3 \, \delta \left( \left({2\pi \over N_c}\right)^3 w_2 w_3 (c-c_0) - z \right) \, \delta \left(\hat{\xi}  \left({2\pi \over N_c} \right)^{3/2} {w_3 \over w_2} - c \right)\\
    =& \left\{
    \begin{array}{l}
     \left({N_c \over 2\pi}\right)^{3} {1\over 2 c (c-c_0)} \quad {\rm for}\ 0\leq z \leq \hat{\xi}^{-1} \left({2 \pi \over N_c}\right)^{3/2} c (c-c_0), \ c \leq \hat{\xi} \left({2 \pi \over N_c}\right)^{3/2}, \\
     \left({N_c \over 2\pi}\right)^{3} {1\over 2 c (c-c_0)} \quad {\rm for}\ 0\leq z \leq \hat{\xi} \left({2 \pi \over N_c}\right)^{9/2} {c-c_0 \over c} , \ c \geq \hat{\xi} \left({2 \pi \over N_c}\right)^{3/2}.
    \end{array}
    \right.
   \end{split}
  \end{equation}
  Since the region of our interest is given by $c \geq c_0 \sim 12.2 > \hat{\xi} (2\pi / N_c)^{3/2}$ (assuming $\hat{\xi} < 1$, $N_c > 1$), we can perform the integration with respect to $c$ (see the left hand side of Figure \ref{cz-plane} for an illustration):
  \begin{equation}
   \begin{split}
    P(z) =& N_0^{-1} \int_{{c_0 \over 1-z  \hat{\xi}^{-1} ({N_c / 2\pi})^{9/2} }}^{c_1} dc \, P(z,c)\\
    =& \hat{\xi}^{-1} \left({N_c \over 2\pi}\right)^{9/2} {1\over b_1} \ln \left[\hat{\xi} \left({2\pi \over N_c}\right)^{9/2} {b_1 \over  z}\right], \quad 0 \leq z \leq \hat{\xi} \left( {2\pi \over N_c}\right)^{9/2} {b_1},\\
    N_0 =& \int_0^{\hat{\xi} \left(2 \pi \over N_c\right)^{9/2} b_1} dz  \int_{c_0 \over 1-z  \hat{\xi}^{-1} ({N_c / 2\pi})^{9/2}}^{c_1} dc \, P(z,c) = \hat{\xi} \left(2\pi \over N_c\right)^{3/2}  {b_1 \over 2 c_0},\\
    b_1 =& {c_1 - c_0 \over c_1},
   \end{split}
   \label{z=w_1w_2(c-c_0) without xi}
  \end{equation}
  where $N_0$ is the normalization constant. If we input the value obtained in (\ref{condition for C}) for $c_1\sim 13.0$, we have $b_1  \sim 0.0610$.

  If we recover the coefficient appearing in (\ref{lambda with coefficient}), the probability distribution for $\hat{\Lambda}$ can be read off by using a relation:
  \begin{equation}
   P\left({\hat{\Lambda}}\right) \, d\left({\hat{\Lambda}}\right) = P(z) \, dz.
    \label{converting to Lambda}
  \end{equation}
  Now the product distribution for $\hat{\Lambda}$ is written by
  \begin{equation}
   P\left({\hat{\Lambda}}\right) = {2500 \sqrt{5} \over 3b_1} \hat{\xi}^{-1} \left({N_c \over 2\pi}\right)^{9/2} \ln \left[{3 \over 2500 \sqrt{5}} \hat{\xi} \left({2\pi \over N_c}\right)^{9/2} {b_1 \over \hat{\Lambda}}\right].
    \label{PDF of simplest Lambda}
  \end{equation}
  So we see that the probability distribution for $\hat{\Lambda}$ peaks at $\hat{\Lambda}=0$ with a divergent behavior $\ln 1/\hat{\Lambda}$. 
  
  Note that we have ignored that $\Lambda$ can be negative. The probability for negative $\Lambda$ is actually larger than that for positive $\Lambda$. It is easy to extend the calculation to include it. However, we assume that the universe starts in the inflationary epoch with a relatively large vacuum energy density so it will reach the small positive $\Lambda$ region before the negative region. This allows us to simply focus on the peaking behavior of $P(\Lambda)$.

  Next we consider the expectation value of $\hat{\Lambda}$, using the distribution (\ref{PDF of simplest Lambda}).
  The value for positive $\hat{\Lambda}$ is given by
  \begin{equation}
   \begin{split}
    \left< {\hat{\Lambda}} \right> =& \int_0^{\left({3\over 2500 \sqrt{5}} \right) \hat{\xi} \left(2 \pi \over N_c\right)^{9/2} {b_1}} d \left({\hat{\Lambda}} \right) \, {\hat{\Lambda}}\, P \left({\hat{\Lambda}} \right) \\
    =&{3\over 2500 \sqrt{5}} \hat{\xi}  \left({2\pi \over N_c}\right)^{9/2} {b_1 \over 4}.
   \end{split}
  \end{equation}
  Plugging the values in (\ref{condition for C}), we get
  \begin{equation}
   \begin{split}
    \left< {\hat{\Lambda}} \right> \sim & 8.19 \times 10^{-6}  \hat{\xi} \left(2 \pi \over N_c\right)^{9/2}.
   \end{split}
   \label{expectation value for simple LVS}
  \end{equation}
 where  $\hat \xi$ in (\ref{LVS potential}) suggests that $\hat \xi \sim 10^{-3}$ to $10^{-2}$ is not unreasonable. We can also calculate the variance of this system by
  \begin{equation}
   \begin{split}
    \left< {\hat{\Lambda}}^2 \right> =&\left({3\over 2500 \sqrt{5}}\right)^2 \hat{\xi}^2 \left({2\pi \over N_c}\right)^{9} {b_1^2 \over 9}
    \sim 1.19 \times 10^{-10} \hat{\xi}^2 \left({2\pi \over N_c}\right)^{9}.
   \end{split}
   \label{variance for simple LVS}
  \end{equation}
  Although the ratio $2\pi/N_c$ might make the quantities much smaller as a fine-tuning, we can conclude that the expectation value (\ref{expectation value for simple LVS}) has the typical scale $10^{-5}$ which is the outcome of moduli stabilization (dynamics) of the potential.
  It is interesting that the value obtained here is close to the constrained value of the bracket in the potential (\ref{constraint for the bracket part}) though we take into account the probability distribution of each parameters.
  This is because the model we considered include just two random variables as inputs.
  Once there appear multiple variables product in the coefficient of the entire potential, we can easily expect that such the many random variables contribute to make the cosmological constant smaller.

  Since we linearize the function for $\hat \Lambda (c)$ around the vicinity of $\hat{\Lambda} = 0$ to $(c-c_0)$, $P(\hat \Lambda)$ (\ref{PDF of simplest Lambda}) is a good approximation for relatively small $\hat \Lambda$.
 Note that here we apply the upper bound $c_1$ given by (\ref{condition for C}) for convenience.
 If we compare the potential (\ref{Westphal's potential}) and the approximated potential (\ref{Westphal potential at the vicinity of zero}), the difference in $\Lambda$ away from zero may differ a bit, but will not change the essential behavior around zero.

 \subsection{A measure on the suppression of $\Lambda$}
 
For our application to the $\Lambda$ problem in string theory, we like to introduce a measure on how much the moduli stabilization dynamics will suppress the value of $\Lambda$.
For $\Lambda=\Pi_{j=1}^N x_j^{n_j}$, let
 \begin{equation}
  Q = \frac{1}{2} \log \left[\frac{\left< x^2_1 \right>^{n_1}\left< x^2_2 \right>^{n_2}\left< x^2_3 \right>^{n_3} \cdots \left< x^2_N \right>^{n_N}}{\left< \Lambda^2  \right>} \right]
 \end{equation}
 where $Q$ measures the order of magnitude the average magnitude of $\Lambda$ is suppressed relative to its naive expectation, and we use $\log$ with base 10.
 We consider $\sqrt{\left< x_i^2 \right>}$ instead of $\left< x_i \right>$ so that the divergent behavior of negative cosmological constant toward $\Lambda = 0$ is also taken into account to argue the preference of small magnitude of $\Lambda$.
 \footnote{Here we consider only the vicinity of $\Lambda=0$. In this region, we may expect that the probability distribution of negative $\Lambda$ has approximately same behavior as that of positive $\Lambda$, even though there is a concern about BF bound, above which some tachyonic modes are allowed.}
 For example, (\ref{pMoments}) yields $Q=0$ for $\Lambda=x_1 x_2 \cdots x_n$ where $x_i$ are independent random variables.
 In the absence of moduli stabilization, $Q \le 0$ generically. In the presence of moduli stabilization, $\Lambda$ is meaningfully given only at the meta-stable minimum of the effective potential. Both the extremum and stability conditions introduce constraints., which can enhance the value of $Q$. 
 
Applying to the problem at hand, we have, referring to (\ref{lambda with coefficient}),
 \begin{equation}
  Q = \frac{1}{2} \log \left[\frac{\left<w^2_1\right>^3\left<w^2_2\right>\left<w^2_3\right>\left<(c-c_0)^2\right>}{\left<w^6_1w^2_2w^2_3(c-c_0)^2\right>} \right]
   \label{measure of order of smallness}
  \end{equation}
 where the expectation values in the numerator are evaluated before imposing the constraints, while that in the denominator is evaluated after imposing the constraints.
 
 Let us consider the potential (\ref{Westphal's potential}) before solving for $V_{\rm min}$. Apriori, the bracket involving $C$ and $x$ can take any value without bound. However, the approximation for a valid $V$ probably breaks down when $V/M_P^4$ exceeds unity. So let us simply assign the naive expected value to be of order unity. On the other hand, we find that the actual expected value is given in (\ref{variance for simple LVS}),
 \begin{equation}
  \sqrt{\left< \hat{\Lambda}^2  \right>} \sim 1.09 \times 10^{-5}  \hat{\xi} \left(2 \pi \over N_c\right)^{9/2}
  \label{expected value in simple case}
 \end{equation}
 If there are no constraints for both extrema and stability, the potential value is expected of order of Planck scale because the system we consider here breaks down above Planck scale.
 Then the numerator in the log function of (\ref{measure of order of smallness}) is expected to be about $1$ in Planck unit.
 Thus we see that $Q \sim  4.96 - \log \hat{\xi} + 9/2 \log (N_c/2 \pi)$, which means that the string dynamics in the single modulus case can easily reduce $\Lambda$ by five orders of magnitude or more.

 Although so far we consider up to only four random parameters, these random parameters themselves may be non-trivial functions of the other stabilized moduli in flux compactification. 
 Following the argument around (\ref{multiple vs. product in vev}), we see that additional random parameters (rather than additional powers of the same random parameters) may suppress the expected value (\ref{expected value in simple case}) further.

\section{Discussions and Remarks}\label{Remarks}

In this paper, we study a single modulus K\"ahler uplifting model in the large volume scenario within Type IIB flux compactification. 
We study the probability distribution of classically stabilized vacua and show the emergence of the peaking (at zero) behavior for a small $\Lambda$
\footnote{A power distribution was applied in a {\it priori} distribution for the cosmological constant \cite{Garriga:1999bf} in the context of anthropic principle \cite{Weinberg:1987dv}.
However since a {\it priori} distribution was considered to come from the inflationary model, a {\it priori} distribution of the vacuum energy $P\left(V(\phi)\right) \propto 1/|V'(\phi)|$ is mostly constant due to slow-roll conditions for inflation \cite{Weinberg:2000qm,Weinberg:2000yb,Garriga:2000cv}.}.
However, in the absence of fine tuning, the smallness of the expected $\Lambda$ is still far too big to explain the observed dark energy.
On the other hand, we see that the above estimates of $\Lambda$ assume that the parameters entering $V$ (such as $W_0$ and $A_1=-A$ in the superpotential $W$) have uniform or normal distributions. This is a very conservative estimate.

It is simplest and natural to have $W_0=A_1=0$, yielding supersymmetric Minkowski solutions. So we do expect zeros for them to be very likely. Since other values are also possible, as we have pointed out, we believe that their probability distributions naturally peak at zero values.
In actual cases, they are functions of the stabilized heavy modes, so they themselves can be products of other random variables. In the simple examples considered  for $W_0$ (\ref{zeroW0}) and $A_1$ (\ref{zeroA1}), we see that this is entirely plausible. That is, by the time they enter into $V$, their probability distributions can be very peaked (at zero) already. This will in general lead to a much more peaked behavior (at zero) in the probability distribution of $\Lambda$.
In general, we can check this key point by considering multi-moduli models and stabilize the moduli one by one and check the behavior of the probability distributions of the output parameters. 

It remains to be seen whether the above mechanism is one of the main reasons underlying the very small observed $\Lambda$. If that is the case, the smallness of $W_0A$ in (\ref{Leg})
(or a similar factor in a more realistic multi-moduli potential)
renders the potential to be very small and almost flat around the minimum of $V$, so it will also render the modulus mass to be exponentially small.
When the modulus mass is light enough, we may worry about the cosmological moduli problem \cite{Banks:1993en,deCarlos:1993jw}.
The cosmological moduli problem is absent when $m_{t_1} \gtrsim {\cal O} (10) \, {\rm TeV}$.
If we naively estimate the physical modulus mass to be $m_{t_1} \sim \sqrt{\left<\hat{\Lambda} \right> } M_P$
\footnote{The physical modulus mass is estimated to be $m^2_{t_1} \sim K^{T \bar{T}} \partial_{t_1}^2 V = M_P^2 4x^2/3 \, \partial_x^2 (V/M_P^4)$.
Since we expect $x \sim 3$ and $\left< \partial_x^2 (V/M_P^4)\right> \sim \left< \hat{\Lambda} \right>$ in stable positive vacua, we roughly estimate $m_{t_1} \sim \sqrt{\left<\hat{\Lambda} \right> } M_P$.}
, we can safely apply the mechanism of product distribution for smaller cosmological constant down to $\hat{\Lambda} \gtrsim 10^{-30}$, which is still far from the observed value of $10^{-123}$. 
 It remains to be seen whether the situation will improve in a more realistic multi-moduli model.

In the above single modulus model, it is easy for us to see that the stability condition is not too restrictive. That is, among the cases where the extremum condition is satisfied, a reasonable fraction ${\cal P}$ (about a few percent) of the extremum solutions also satisfies the stability condition.
As we introduce more moduli into the scenario, we expect that this ratio ${\cal P}$ will become Gaussianly suppressed \cite{Aazami:2005jf,Chen:2011ac,Marsh:2011aa} (see also \cite{Denef:2004cf}).
This is simply a result of (mass squared) eigenvalue repulsion of the Hessian matrix $\partial_i \partial_j V$ of the potential $V$. So the following picture emerges. At relatively high vacuum energy densities, there are very few de-Sitter vacua (if any at all) in some regions of the landscape (the regions under analysis), because the number of stable vacua is Gaussianly suppressed. At lower energies, the number of low scale moduli involved decreases, so the
probability of stability solutions is no longer Gaussianly suppressed. This multi-moduli picture is consistent with the single modulus analysis carried out in this paper; how these two pictures merge together will be the challenge for the future.

\section*{Acknowledgment}
We have benefited from fruitful discussions with Xingang Chen, Michael Haack, Liam McAllister, Gary Shiu, Alexander Westphal and Timm Wrase.
SHHT is supported in part by the National Science Foundation under grant PHY-0355005.

\appendix

\section{Other scenarios for the parameters in $V$ of the single modulus model \label{sec:other-scen-param}}
 
  In this appendix, we consider three more plausible scenarios for the probability distribution of $\Lambda$ (\ref{constrained product distribution from Westphal model}).

   \paragraph{Case 1}
  Although we neglected the randomness of $\hat{\xi}$ in (\ref{simple random model from westphal}) because of its discrete behavior around zero value, it is possible that we can treat this parameter as a continuous variable passing through zero after including stringy loop corrections in the K\"ahler potential.
  The stringy loop corrections in the K\"ahler potential are studied in \cite{Berg:2005ja,Berg:2004ek,Cicoli:2007xp,Cicoli:2008va,Berg:2007wt} (see also supergravity loop correction in \cite{vonGersdorff:2005bf}).
  The typical effect of stringy loop correction can be written down \cite{Berg:2007wt},
  \begin{equation}
   \begin{split}
    {K \over M_P^2} = -2 \ln {\cal V} - {\hat{\xi} \over {\cal V}} + \sum_{a} {g_K^a {\cal E}_a^{(K)} g_s \over {\cal V}} + \sum_q {g_W^q {\cal E}_q^{(W)} \over {\cal V}},
   \end{split}
  \end{equation}
  where we expand the K\"ahler potential (\ref{LVS potential}) up to leading order of ${\hat{\xi}}$
  \footnote{The $\hat{\xi}$ parameter itself is also affected by the loop correction to be $\hat{\xi} \rightarrow \hat{\xi} (1+\pi^2 g_s^2 /3\zeta(3))$ (see review \cite{Green:1999qt}), but we neglect this effect here because this correction doesn't help to have smooth behavior around $\hat{\xi} = 0$.}.
  The terms related to ${\cal E}_a^{(K)}$ are from the KK-modes, while terms with ${\cal E}_q^{(W)}$ originate from the exchange of winding strings.
  They are functions of complex structure moduli, and therefore given as constants after stabilization.
  The $g_K^a$ is proportional to some two-cycle volume, while the $g_W^q$ is inversely proportional to some two-cycle volume.

  Though these stringy loop corrections depend on the K\"ahler moduli, they are sub-leading terms in general and will not change the stabilization qualitatively under the assumption of large volume.
  If we deal with the string loop corrections as effective constants as well as $\hat{\xi}$ parameter for simplicity, we may treat the combined parameter among $\alpha'$ and stringy loop corrections appearing in K\"ahler potential as a random variable passing through zero value.

\begin{figure}
 \begin{center}
  \includegraphics[width=19em]{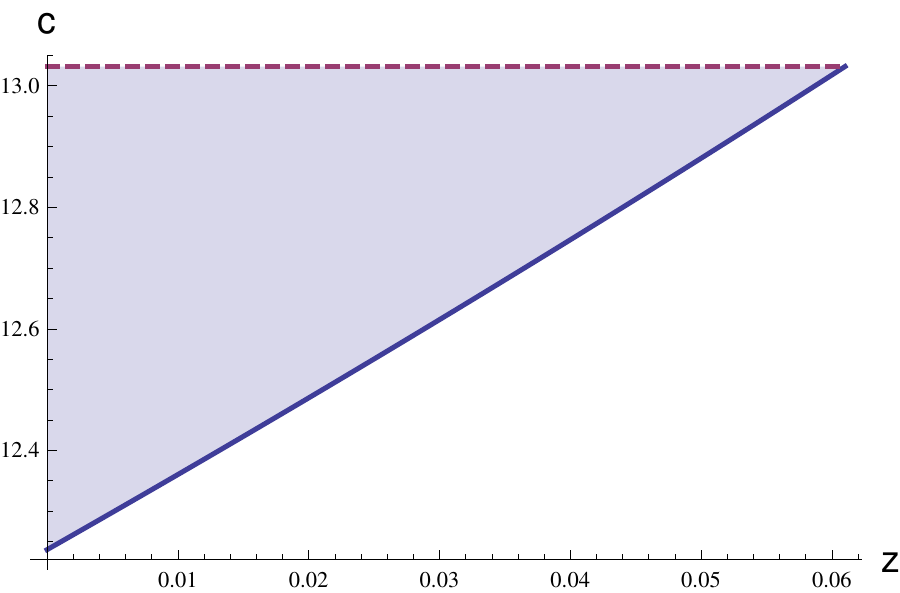}
  \hspace{1em}
  \includegraphics[width=19em]{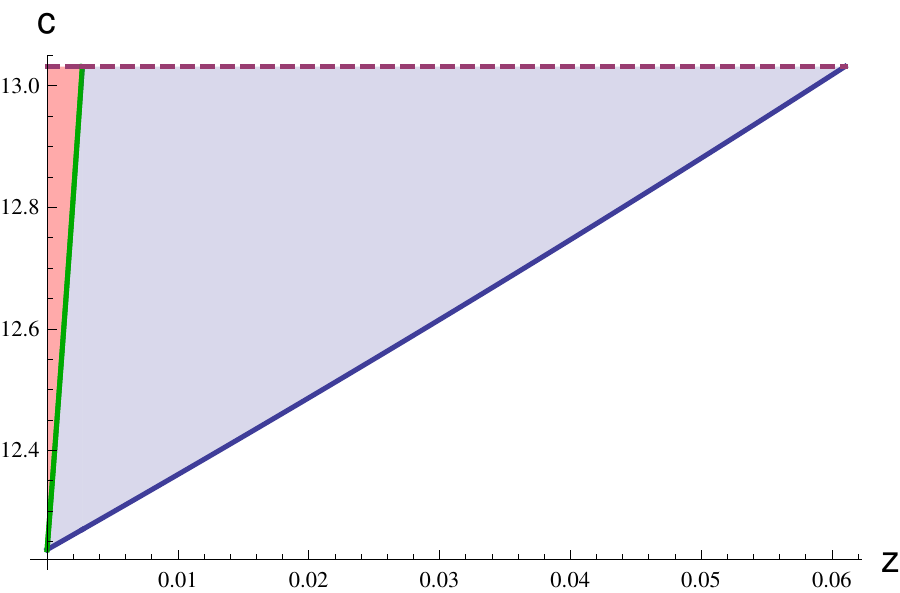}
 \end{center}
 \caption{\footnotesize The regions that contribute to $P(z,c)$ in the the $c$-$z$ plane.
 The left plot shows the valid region in the cases of (\ref{simple random model from westphal}), (\ref{constrained product distribution with xi}), and (\ref{constrained product distribution without xi}), where the region is surrounded by $c=c_1$ and $c = c_0/(1-z)$.
 The right plot is for (\ref{including peaking in a_1}), where the middle curve $c=1/(1-y_m^{9/2} z)$ which separates two regions corresponding each function in (\ref{cz distribution for the most complicated one}).
 Here we assume $\hat{\xi}=1, a_1=1, \gamma_1 = 1$ in the figures.}
 \label{cz-plane}
\end{figure}

  Neglecting $w_1$ and recovering $w_4$ dependence, the system of interest now becomes,
  \begin{equation}
   z = \left( 2\pi \over N_c \right)^3 w_2 w_3 (c-c_0), \quad c = \left( 2\pi \over N_c \right)^{3/2} {  w_3 w_4 \over w_2}, \quad c_0 \leq c \leq c_1.
    \label{constrained product distribution with xi}
  \end{equation}
  where $w_j$ have uniform distributions $0 \le w_j \le 1$.
  Performing integral with constraints in the same way as that of (\ref{z=w_1w_2(c-c_0) without xi}) (the valid region is illustrated in the left hand side of Figure \ref{cz-plane}), we get, after some calculations, 
  \begin{equation}
   P(z) = \left( {N_c \over 2\pi} \right)^{9/2} {2\over b_1} \left[\left( {N_c \over 2\pi} \right)^{9/2} {z \over b_1} + \ln \left(\left( {2\pi \over N_c} \right)^{9/2}{b_1 \over z}\right) -1 \right], \quad
    0 \leq  z \leq \left( {2\pi \over N_c}\right)^{9/2} {b_1},
  \end{equation}
  where $b_1 = (c_1 - c_0)/c_1$.
  Using (\ref{converting to Lambda}), we can convert to $P(\hat{\Lambda})$ ($\hat{\Lambda} = \Lambda/M_P^4$).
  The expectation value and the variance in this case become
  \begin{equation}
   \begin{split}
    \left< {\hat{\Lambda}} \right> =& {3\over 2500 \sqrt{5}} \left({2\pi \over N_c}\right)^{9/2} {b_1 \over 6}
    \sim 5.46 \times 10^{-6} \left({2\pi \over N_c}\right)^{9/2},\\
    \left< {\hat{\Lambda}}^2 \right> =&\left({3\over 2500 \sqrt{5}}\right)^2 \left({2\pi \over N_c}\right)^{9} {b_1^2 \over 18}
    \sim 5.96 \times 10^{-11} \left({2\pi \over N_c}\right)^{9},
   \end{split}
  \end{equation}
  for positive $\Lambda$.

  \paragraph{Case 2}
  Next, let us consider the model without taking $\hat{\xi}\, ( = w_4)$ random, but recovering $a_1\, (w_1)$ as a random variable. 
  The system of our interest now becomes
\begin{equation}
 z = w_1^m w_2 w_3 (c-c_0), \quad c = \hat{\xi} {w_1^n w_3 \over w_2}, \quad c_0 \leq c \leq c_1.
  \label{constrained product distribution without xi}
\end{equation}
with $m=3, n=3/2$.
Let us first derive the analytic formula with arbitral $m,n $ satisfying $m > n$.
We assume that $w_1, w_2, w_3$ obey uniform distribution from 0 to 1 just for simplicity.
The product distribution for fixed $z,c$ are given by an integration:
\begin{equation}
 \begin{split}
  P(z,c) =& \int \, dw_1 dw_2 dw_3 \, \delta \left( w_1^m w_2 w_3 (c-c_0) - z \right)\, \delta \left(\hat{\xi} {w_1^n w_3 \over w_2} - c \right)\\
  =& \left\{
  \begin{array}{l}
   {1\over 2(m-1)} \left[ \hat{\xi}^{1-m \over m-n} \left(c(c-c_0)\right)^{n-1 \over m-n} z^{1-m \over m-n} - {1\over c(c-c_0)}\right] \ {\rm for} \ \hat{\xi}^{-m/n} c^{m/n} (c-c_0) \leq z \leq \hat{\xi}^{-1} c(c-c_0), \\
   {1\over 2(m-1)} \left[ \hat{\xi}^{m-1 \over m+n} c^{{n-1 \over m-n} - {2m(m-1)\over m^2-n^2}} (c-c_0)^{-{n+1 \over m+n}} z^{1-m \over m+n} - {1\over c(c-c_0)} \right] \quad {\rm for} \ z\leq \hat{\xi} {c-c_0 \over c}, \hat{\xi} \leq c.
  \end{array}
  \right.
 \end{split}
\end{equation}
The remaining task is to obtain the probability distribution $P(z)$ after integrating over $c$ in the appropriate regions.
Now let us use the actual numbers of interest to us here: $m=3, n=3/2, c_0 = 200 \sqrt{5} e^{-5/2} /3 \sim 12.2 > \hat{\xi}$.
The appropriate region in the $c$-$z$ plane is illustrated in the left hand side of Figure \ref{cz-plane}.
The probability distribution function in this case now becomes
\begin{equation}
 \begin{split}
  P(z) =& N_0^{-1}\int_{c_0 \over 1- \hat{\xi}^{-1} z}^{c_1} dc \, {1\over 4} \left[ \hat{\xi}^{9/4} c^{-13/9} (c-c_0)^{-5/9} z^{-4/9} - {1\over c(c-c_0)} \right]\\
  =& \hat{\xi}^{-1} {5 \over 16 b_1} \left[9 \left({b_1 \over \hat{\xi}^{-1} z }\right)^{4/9} + 4 \ln \left({\hat{\xi}^{-1} z \over b_1}\right) -9 \right],\\
  N_0 =& \int_0^{\hat{\xi} {b_1}} dz  \int_{{c_0 \over 1-\hat{\xi}^{-1} z}}^{c_1} dc \, P(z,c) = \hat{\xi} {b_1 \over 5 c_0},
  \label{case2P}
 \end{split}
\end{equation}
where $N_0$ is the normalization constant.
The $P(z)$ for the system (\ref{constrained product distribution without xi}) is diverging as $z^{-4/9}$. In the absence of the constraint, Table \ref{tab:asymptotics of product distribution} tells us that $P(z) \sim z^{-1+1/3}=z^{-2/3}$, so we see that the constraint reduces the divergent power by $2/9$.
Using this formula (\ref{case2P}), we can calculate the expectation value and variance of $\hat{\Lambda}$,
\begin{equation}
 \begin{split}
    \left< {\hat{\Lambda}} \right> =& {3\over 2500 \sqrt{5}} \hat{\xi} {4 b_1 \over 56}
    \sim 2.34 \times  10^{-6} \hat{\xi},\\
    \left< {\hat{\Lambda}}^2 \right> =&\left({3\over 2500 \sqrt{5}}\right)^2 \hat{\xi}^2 {5 b_1^2 \over 207}
    \sim 2.59 \times 10^{-11} \hat{\xi}^2.
 \end{split}
\end{equation}
Since we treated $w_1 = a_1/\gamma_1^{2/3}$ as a random parameter here, all possible values for $w_1$ are taken into account without fine-tuning.

\paragraph{Case 3}

In the previous case, we assume a uniform distribution for $w_1$.
If we consider only the $SU(N_c)$ gaugino condensation on D7-branes and assume randomness for $N_c$ of $a_1 = 2\pi/ N_c$ with fixed $\gamma_1$, we may expect that probability distribution is more peaked toward $\hat{\Lambda} = 0$.
Here for simplicity, we assume uniform distribution from 0 to 1 for $w_2, w_3$, and uniform distribution from 1 to $y_m\propto N_c^{max}$ for $y_1=1/w_1$, and consider the following constrained product distribution:
\begin{equation}
 \begin{split}
  z =& {w_2 w_3 \over y_1^3} (c-c_0), \quad c = \hat{\xi} {w_3 \over y_1^{3/2} w_2}.
 \end{split}
 \label{including peaking in a_1}
\end{equation}
The distribution for this system can be calculated,
\begin{equation}
 \begin{split}
  P(z,c) =& \int_1^{y_m} dy_1 \int_0^{1} dw_2 \int_0^1 dw_2 \,
  \delta \left( {w_2 w_3 \over y_1^3} (c-c_0) - z \right)\, \delta \left(\hat{\xi} {w_3 \over y_1^{3/2} w_2} - c \right)\\
  =& \left\{
  \begin{array}{l}
   {y_m^4 - 1 \over 8 c (c-c_0)} \quad {\rm for} \ 0\leq z \leq \hat{\xi} {c-c_0 \over y_m^{9/2} c},\\
   {1\over 8} \left[ \hat{\xi}^{8/9} c^{-17/9} (c-c_0)^{-1/9} z^{-8/9} - {1\over c (c-c_0)}\right] \quad {\rm for} \ {c-c_0 \over y_m^{9/2} c} \leq z \leq \hat{\xi} {c-c_0 \over c}.
  \end{array}\right.
 \end{split}
 \label{cz distribution for the most complicated one}
\end{equation}
Now we integrate over $c$ for the appropriate regions illustrated in the right hand side of Figure \ref{cz-plane}.
The resulting probability distribution of $z$ can be written by, after some calculations, 
\begin{equation}
 \begin{split}
  P(z) =& \left\{
  \begin{array}{l}
   \hat{\xi}^{-1} {\sqrt{y_m} \over 64 b_1 (\sqrt{y_m}-1)} \left[ -36 \ln y_m + (y_m^4 - 1) \left(9 + 8 \ln { b_1 \over y_m^{9/2}  \hat{\xi}^{-1} z} \right)  \right] \quad {\rm for}\ 0 \leq z \leq \hat{\xi}  {b_1 \over y_m^{9/2}},\\
    \hat{\xi}^{-1} {\sqrt{y_m} \over 64 (\sqrt{y_m}-1)  \hat{\xi}^{-1} z} \left[9 \left({\hat{\xi}^{-1} z \over b_1}\right)^{1/9} - {\hat{\xi}^{-1} z \over b_1} \left(9 + 8 \ln { b_1 \over  \hat{\xi}^{-1} z}  \right)\right] \quad {\rm for}\  \hat{\xi} {b_1 \over y_m^{9/2}} \leq z \leq \hat{\xi} {b_1}.
  \end{array}
  \right.
 \end{split}
\end{equation}
Applying this distribution to obtain the expectation values, we have
\begin{equation}
 \begin{split}
  \left< {\hat{\Lambda}} \right> =& {3\over 2500 \sqrt{5}} \hat{\xi} {b_1 \over 40 y_m^{9/2}} \left[ 1+ y_m^{1/2} (1+ y_m^{1/2} + y_m)(1+y_m^{3/2} + y_m^3)\right]\\
    \stackrel{y_m \rightarrow \infty}{\rightarrow}& 8.19 \times  10^{-7} \hat{\xi},\\
    \left< {\hat{\Lambda}}^2 \right> =& \left({3\over 2500 \sqrt{5}} \right)^2 \hat{\xi}^2 {b_1^2 \over 171 y_m^9} \left[1 + y_m^{1/2}(1+y_m^{1/2})\left((1+ y_m)^2 - y_m \right)\left((1+ y_m^3)^2 - y_m^3 \right)\right]\\
  \stackrel{y_m \rightarrow \infty}{\rightarrow}& 6.27 \times  10^{-12} \hat{\xi}.
 \end{split}
\end{equation}
where we also give the numerical values at $y_m \rightarrow \infty$.

\bibliographystyle{utphys}
\bibliography{myrefs}

\end{document}